# AI Persona, Service Consumption, and User Intent Entropy: Field Experimental Evidence from an LLM Platform

[1]Junjie Li[a], Xiaofan Li[b], Lauren Xiaoyuan Lu[c], Yiwei Wang[d], Bruce Yang[b,e]

[a] Zhejiang University, School of Management, Zhejiang, China, junjie.li.som@zju.edu.cn

[b] National University of Singapore, School of Computing, Singapore, li.x@nus.edu.sg

[c] Dartmouth College, Tuck School of Business, Hanover, USA, lauren.x.lu@tuck.dartmouth.edu

[d] Zhejiang University, International Business School, Zhejiang, China, yiweiwang@intl.zju.edu.cn

[e] Singapore Sapiens Technology Pte. Ltd., Bruce@sapiens-ai.io

September 2026

**Abstract**

**Problem definition:** Firms deploying large language model (LLM) services must decide not only what their AI can do but also how it communicates with users. We examine how a relational persona, designed to be warmer, more empathetic, and more engaging than a non-relational persona, affects service consumption and how users' service objectives develop through interaction. **Methodology/results:** We conduct a randomized field experiment with 9,586 newly registered users of an LLM platform, holding the underlying model and service capabilities constant. The relational persona increases service interactions (service sessions, +8.1%; cumulative duration, +10.6%; chat rounds, +24.2%; user intent entropy, +5.8%) and service outputs (files, +12.3%; distinct goals, +12.1%). These average effects mask substantial heterogeneity by entry intent. First-session effects are insignificant for *Task Execution* users. *Socialization* and *Knowledge Exploration* users show similar increases in chat rounds, but through different paths: *Socialization* increases intent entropy without more outputs, whereas *Knowledge Exploration* increases outputs without higher intent entropy. Representing intent dynamics as a transition process of intent states, we find that the relational persona increases intent transition entropy for *Socialization* users (+11.8%) but intent continuation probability for *Knowledge Exploration* users (+12.8%), suggesting greater conversational breadth in the former and persistence in the latter. In subsequent use, the relational persona increases aggregate chat rounds and service outputs across all entry intents. Session count rises by 12.2% for *Task Execution* and 49.4% for *Socialization*, but not significantly for *Knowledge Exploration*. Effects on session count and intent entropy strengthen over time, whereas output effects remain stable. **Managerial implications:** AI persona is an operational design lever rather than just a presentation feature. Because more interactions do not uniformly generate more outputs, firms should evaluate interactions and outputs separately and consider matching persona design to user intent, particularly when additional interactions consume costly computing resources.

[1] The authors are listed in alphabetical order.

## 1 Introduction

Large language models (LLMs) are emerging as a new class of digital service systems through which firms deliver services directly to customers. Like conventional self-service technologies (SSTs), they enable customers to obtain services without direct employee assistance (Meuter et al. 2000). Conventional SSTs, such as ATMs and self-ordering kiosks, typically guide customers through predefined activities toward a bounded service objective. LLM services, by contrast, accommodate heterogeneous and potentially underspecified needs through open-ended conversations. Users may clarify, elaborate, revise, or redirect their requests as AI responses introduce new information, with each subsequent user message becoming a new input to service production. Customer participation therefore extends beyond following predefined steps to shaping what service is requested and how it is delivered. The amount of service consumption, the intents pursued, and the resulting outputs can develop through user–AI interaction.

This structure gives firms a new set of operational design choices. Firms must decide not only what their AI can do but also how it should communicate with users. One such choice is the AI's *persona*, a system configuration that governs its tone, vocabulary, interaction strategy, and social expressiveness. A *relational* persona, designed to be warmer, more empathetic, and more engaging than a non-relational persona, may encourage more frequent and extensive service use. Yet additional interactions consume service capacity without necessarily generating more outputs. The operational question is therefore how persona design changes service consumption and when more interactions it encourages are accompanied by greater service outputs.

Existing research provides only partial guidance. Studies of digital service operations show that technology design and customer coproduction can alter channel use, service demand, and firm performance (Xue et al. 2007, Campbell and Frei 2009), while service-process research highlights how the organization and sequence of activities shape customer outcomes (Bellos and Kavadias 2020). Research on AI anthropomorphism further demonstrates that social and conversational cues can affect customer responses, information disclosure, and transaction conversion (Luo et al. 2019, Blut et al. 2021, Schanke et al. 2021, Xu et al. 2024). These studies largely examine individual design cues or immediate outcomes within predefined service processes. We instead examine how AI persona shapes open-ended conversations. Our focus is not simply whether users interact more, but which service intents they pursue, how those intents evolve through conversation, and what service outputs their interactions generate.

We address this gap by asking three questions. First, how does a relational AI persona affect service interactions and service outputs? Second, how do these effects vary with users' entry intents, and what patterns of intent continuation and transition accompany them? Third, how do these effects extend beyond the initial service encounter and evolve with subsequent interactions?

We study these questions through a randomized field experiment on Agnes AI, an LLM platform that operates as a self-service technology without human involvement. The platform accommodates heterogeneous service needs within a single persistent conversation thread and can provide information and advice or generate images, videos, slide decks, and reports. We organize each user's conversation into service sessions, with a new session beginning when at least 30 minutes separate consecutive chat rounds. Each round starts with a user message and includes the AI responses and file outputs before the next user message. The experiment involved 9,586 newly registered users, of whom 4,865 were randomly assigned to a relational persona and 4,721 to a non-relational persona for an 18-day treatment period. The relational persona was warmer, more empathetic, and more engaging, whereas the non-relational persona was more concise, task-oriented, and neutral. Both conditions used the same underlying LLM and service capabilities and differed only in persona design.

We find that the relational persona increases total service consumption over the entire sample period across two dimensions: service interactions (service sessions, +8.1%; cumulative duration, +10.6%; chat rounds, +24.2%; user intent entropy, +5.8%) and service outputs (files, +12.3%; distinct goals, +12.1%). Thus, persona design affects not only how much users interact with the service, but also the variety of those interactions and the outputs they generate.

These average effects, however, mask substantial heterogeneity by user intent. Specifically, *Entry Intent* classifies the user's opening message, whereas *User Intent Entropy* summarizes the distribution of intents across all classified chat rounds in a service session. On the platform, users commonly enter with an initial intent and develop different mixes of intents as the conversation unfolds. We classify each user's first-session entry intent from the opening message, which precedes any exposure to the assigned persona, as *Task Execution*, *Socialization*, or *Knowledge Exploration*. Conducting a heterogeneous analysis of users' first-session service consumption by their entry intent, we find no statistically significant effects for *Task Execution* users. For *Socialization* users, the relational persona increases chat rounds by 16.0% and user intent entropy by 9.2%, but does not significantly increase service outputs. For *Knowledge Exploration* users, chat rounds increase by 15.9%, yet user intent entropy remains unchanged, while files created and distinct goals served increase by 18.7% and 18.9%. The first-session heterogeneous analysis results suggest two patterns: more interactions accompany higher intent entropy for *Socialization* users, whereas it accompanies more file production for *Knowledge Exploration* users.

To examine the process underlying these heterogeneous treatment effects of relational persona, we characterize the users' conversational patterns using *Intent Continuation Probability*, which captures the observed probability that consecutive chat rounds express the same intent, and *Intent Transition Entropy*, which captures the variety of observed intent-transition types. For *Socialization* users, the relational persona increases intent transition entropy by 11.8%. For *Knowledge Exploration* users, intent continuation probability increases by 12.8%, while intent transition entropy remains unchanged. These contrasting patterns provide process evidence for the heterogeneous consumption effects: the relational persona broadens the conversational path for *Socialization* users but reinforces continued engagement with the same intent for *Knowledge Exploration* users, for whom more interactions are accompanied by greater service outputs.

The relational persona increases subsequent-session chat rounds and service outputs across all three entry intents, but whether these gains are accompanied by more frequent service use varies by intent. The number of subsequent sessions following the entry intent of *Task Execution* increases by 12.2%, and those following *Socialization* increase by 49.4%; the corresponding estimate for *Knowledge Exploration* is not statistically significant. Later in the treatment period, the relational persona has larger effects on session counts and intent diversity, but a smaller effect on cumulative duration. Its effects on service outputs do not increase significantly over time.

Our study makes three contributions. First, we contribute to research on AI-enabled digital service operations by identifying persona as a key operational design lever that shapes service interactions and outputs while holding the underlying model and capabilities constant. We show that more interactions do not necessarily translate into greater outputs and that these effects vary across service intents and over time.

Second, we extend research on anthropomorphism and human–AI interaction from immediate user responses to the consumption of an *open-ended*, *unstructured* LLM chatbot service. We show that the effects of a relational persona depend on users' entry intents and persist into subsequent interactions, with different patterns for *Socialization*, *Knowledge Exploration*, and *Task Execution* sessions.

Third, we advance research on self-service technology and customer journeys by developing an intent-based view of conversational service processes. Using intent entropy, intent continuation probability, and intent transition entropy, we show that similar increases in interaction volume can follow different paths: relational persona increases intent transition entropy in *Socialization* sessions but induces higher intent continuation probability in *Knowledge Exploration* sessions. This shifts attention from how much users interact to how their service intents evolve through service interactions.

These findings have direct implications for firms deploying LLM chatbot services. Persona should be treated as part of the service operating system rather than merely as a presentation feature. Because its effects vary across users' service intents, applying the same persona universally may be suboptimal. A relational persona can substantially expand subsequent interactions for *Socialization* users, generate more interactions and outputs for *Knowledge Exploration* users, and have limited immediate effects on *Task Execution* users. Firms should evaluate service interactions and service outputs separately. More conversations may involve a broader mix of intents or continued interaction within the same intent, without necessarily generating more outputs. Together, these patterns point toward cost-conscious persona design: because additional interactions consume computing resources whether or not they yield outputs, the persona that maximizes engagement is not always the one that maximizes value.

## 2 Literature Review

Our study draws on three related streams of research. Research on AI-enabled digital service operations examines how AI changes service delivery and operational performance. Research on anthropomorphism and human–AI interaction examines how AI communication design shapes user responses and behavior. Research on self-service technology and customer journeys explains how customer participation and evolving interactions shape service production. Together, these perspectives motivate our study of how AI persona design affects service interactions and outputs, and how user intents develop throughout the service process.

### 2.1 AI-Enabled Digital Service Operations

Large language models (LLMs) are changing how firms deliver digital services, with applications in healthcare, education, customer service, and knowledge work (Noy and Zhang 2023, Brynjolfsson et al. 2025, Bastani et al. 2025, Heinz et al. 2025, Balakrishnan et al. 2025, Wang et al. 2026). Prior research shows that generative AI can improve worker productivity and task performance, but that its effects depend on service design. For example, AI assistance can improve current performance while reducing later unassisted performance (Bastani et al. 2025), while agentic AI can shorten customer-service interactions but lower customer ratings for AI-eligible requests (Wang et al. 2026).

Related work shows that AI design can also affect user participation in digital services. Zhang et al. (2026) find that Google AI Overviews increase engagement in eligible Reddit communities, whereas the later introduction of conversational AI Mode largely removes these gains. This evidence suggests that AI design can affect not only how efficiently a service is delivered, but also how much users consume and interact with the service.

We extend this literature by examining persona as an operational design choice within an LLM service. In a randomized field experiment, we vary the relational character of the chatbot while holding the underlying model and service capabilities constant. We distinguish two dimensions of service consumption: service interactions and service outputs. This distinction allows us to show that greater interaction does not necessarily translate into greater output and that the effects of persona design vary across users' service intents and over time. Our study therefore broadens the operational consequences of AI design from task performance and engagement to how service consumption develops within and across service encounters.

### 2.2 Anthropomorphism and Human–AI Interaction

Our work is also related to research on anthropomorphism in marketing and operations management. Anthropomorphism refers to attributing human characteristics or social capacities to nonhuman entities (Xu et al. 2024). In AI services, firms can create such perceptions through human identities, visual representations, conversational styles, and socially expressive behaviors (Blut et al. 2021). In chatbot settings, conversational design is particularly important because it can shape both user behavior and AI responses (Go and Sundar 2019, Bergner et al. 2023, Ibrahim et al. 2026).

Prior field studies show that anthropomorphic features can affect customer responses, interaction length, information disclosure, and service outcomes (Luo et al. 2019, Schanke et al. 2021, Xu et al. 2024, Fang et al. 2026). Most of these studies, however, examine particular anthropomorphic cues or immediate responses in relatively bounded service settings. Less is known about an integrated persona that operates throughout an open-ended service process in which users pursue heterogeneous objectives and may return repeatedly over time.

We study such an integrated relational persona on an open-ended LLM platform. We show that its effects depend systematically on users' entry intent and persist into subsequent service consumption. Moreover, similar increases in interaction can correspond to different outcomes across service intents. *Socialization* is characterized by broader and more varied service intents, whereas *Knowledge Exploration* is characterized by greater continuation within the same intent and greater service output. Our study therefore extends research on human-like AI design from immediate user responses to how persona shapes the amount, form, and development of service consumption.

### 2.3 Self-Service Technology and Customer Journey

Lastly, our study relates to the self-service technology (SST) literature and its connection to the customer journey. SSTs enable customers to produce services without direct assistance from employees (Meuter et al. 2000). Because customers contribute information, effort, and decisions to service production, their participation can affect both channel usage and firm performance (Xue et al. 2007). SSTs can also expand service consumption. For example, digital banking shifts transactions across channels while increasing overall activity, making its cost and profit effects less clear (Campbell and Frei 2009, Xue et al. 2011). These findings highlight the importance of understanding both how customers participate in service production and how much service they consume.

LLM platforms extend this logic because customer participation is more open-ended. Conventional SSTs, such as ATMs and self-ordering kiosks, typically guide customers through predefined activities toward a bounded service objective. An LLM platform instead accommodates heterogeneous and potentially underspecified requests through conversation, allowing users to clarify an initial intent, elaborate on an AI response, revise their objective, or introduce a new need. Because each AI response can shape the next user input, the service process can evolve dynamically rather than follow a predefined path. The customer-journey perspective provides a useful framework for understanding this variation by examining how service activities develop across interactions and encounters (Lemon and Verhoef 2016, Bellos and Kavadias 2020). Prior research shows that the sequence and configuration of service activities can shape customer responses and outcomes (Dixon and Verma 2013, Dixon et al. 2017). In an LLM service, this evolving service path is reflected in users' service intents as they continue with an existing objective or move among *Task Execution*, *Socialization*, and *Knowledge Exploration* during the conversation.

Building on this perspective, we develop an intent-based approach to characterize these open-ended service processes. Drawing on information entropy and research on conversational state transitions (Shannon 1948, Cappella 1980, Liao et al. 2023), we use *user intent entropy* to capture the diversity and relative prevalence of service intents. We then represent user intent dynamics as a transition process of intent states and characterize *intent continuation probability* and *intent transition entropy*. Together, these measures distinguish interactions that continue with the same service intent from those that develop across a wider variety of intents. This framework allows us to move beyond how much interaction occurs and characterize how users' service objectives develop through interaction, helping explain why similar increases in interaction volume can be accompanied by different service consumption outcomes.

## 3 Institutional Setting and Experiment Setup

### 3.1 Company Background

We conduct our study on Agnes AI, a Singapore-based AI platform founded in April 2025 (Sapiens AI 2025). The platform is powered by Agnes-SeaLLM-8B, an LLM developed by Agnes with approximately 8 billion parameters.[2] Similar to ChatGPT and Claude, Agnes allows users to interact with AI through a chat interface available on mobile devices and the web. Within four months of its public launch in July 2025, the platform reached 150,000 daily active users across Asia, the Middle East, Europe, and North America (Agnes AI 2025). Its large and geographically diverse user base offers a real-world setting for examining how a relational AI persona affects service consumption.

### 3.2 User–AI Service Interactions

Our study examines digital services delivered through the online chat interface of Agnes AI. The platform operated as a self-service technology in which users interact directly with an LLM-enabled chatbot, without the involvement of a human service agent. The chatbot therefore serves as the frontline service provider, communicating with users in natural language and coproducing service outcomes with them. Depending on the request, the chatbot could provide information, offer advice, or generate four types of outputs: images, videos, slide decks, and reports.

At the time of the study, the platform supported a single persistent conversation thread for each user. All exchanges between a user and the chatbot accumulated sequentially within this thread, and users could not open additional threads. Within the single continuous thread, a user engaged in multiple service sessions over time, separated by periods of inactivity. Each session could include multiple chat rounds, with each round consisting of a user message, as well as subsequent chatbot messages and file outputs until the next user message.

[2] The LLM used for the study is *Agnes-SeaLLM-8B*, which is fine-tuned from Qwen3-8B-Base and contains 8.19 billion parameters. It uses a dense decoder-only Transformer architecture with 36 layers, a hidden dimension of 4,096, 32 attention heads, eight key-value heads, and a maximum context length of 40,960 tokens. The model supports English, Chinese, Indonesian, Thai, and Vietnamese and is optimized for Southeast Asian languages, multi-turn dialogue, instruction following, mathematical reasoning, and translation (Hugging Face 2026).

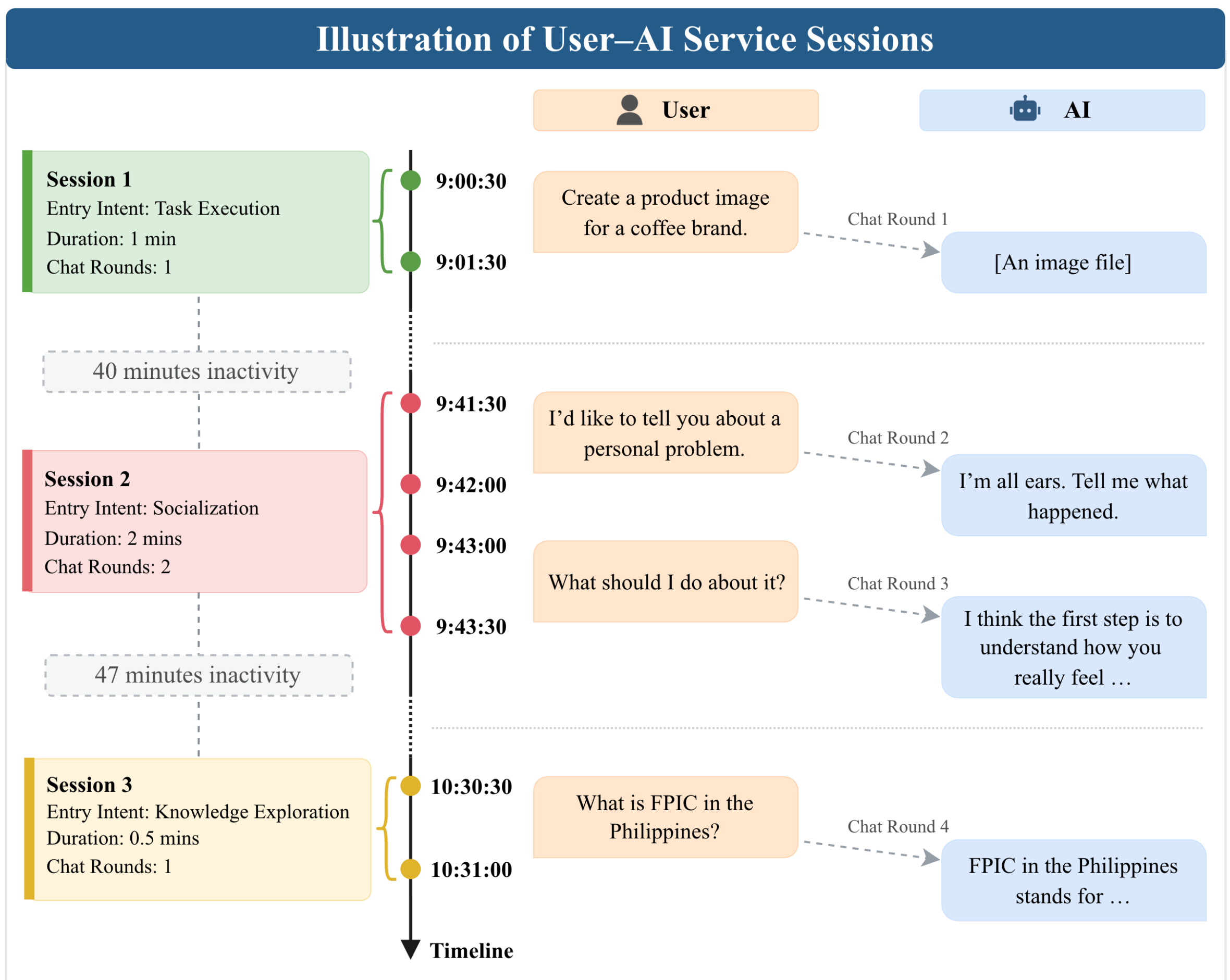

*Notes.* Figure 1 illustrates how a single persistent conversation thread may contain multiple service sessions separated by periods of user inactivity. The sessions may differ in user entry intent and contain different numbers of chat rounds. A chat round consists of a user message and the corresponding AI response.

**Figure 1. Illustration of User–AI Service Interactions**

Figure 1 illustrates this interaction structure using three service sessions within a single conversation thread. In the first session, the user asks the chatbot to create a product image for a coffee brand, and the chatbot delivers the image in a single chat round. After 40 minutes of inactivity, the user returns to the same thread to discuss a personal problem and seek advice. This *Socialization* session contains two chat rounds. Following another 47 minutes of inactivity, the user begins a *Knowledge Exploration* session by asking about FPIC in the Philippines. The example shows how the same user may bring substantially different service needs to the chatbot at different time points.

### 3.3 AI Personas

Agnes AI used persona design to shape how the chatbot interacted with users. The personas were implemented through prompt engineering: persona-specific system prompts guided the generation of every chatbot response, while the parameters of the underlying LLM remained unchanged. The experiment compared two personas that differed in their degree of social expressiveness: a *relational persona* and a *non-relational persona*. The two personas differed along five social and communicative dimensions: *core identity*, *tone and voice*, *vocabulary and syntax*, *interaction strategy*, and *conflict resolution*. The relational persona was designed to be warmer, more empathetic, and more engaging, whereas the non-relational persona was designed to be more concise, task-oriented, and neutral. Appendix B presents the complete system prompts for both personas.

### 3.4 Experiment Design

In December 2025, collaborating with Agnes AI, we conducted a randomized field experiment involving newly registered users. During the six-day pretreatment period (12/6/2025–12/12/2025), the platform identified users eligible for the field experiment. On December 12, the platform randomly assigned 9,586 users, based on their user IDs, to one of the two persona conditions with equal probability. As shown in Figure 2, 4,865 users (50.8%) were assigned to the treatment group and interacted with the relational persona, whereas 4,721 users (49.2%) were assigned to the control group and interacted with the non-relational persona. The assigned persona remained fixed throughout the subsequent eighteen-day treatment period (12/12/2025–12/30/2025). The experiment concluded on December 30, 2025.

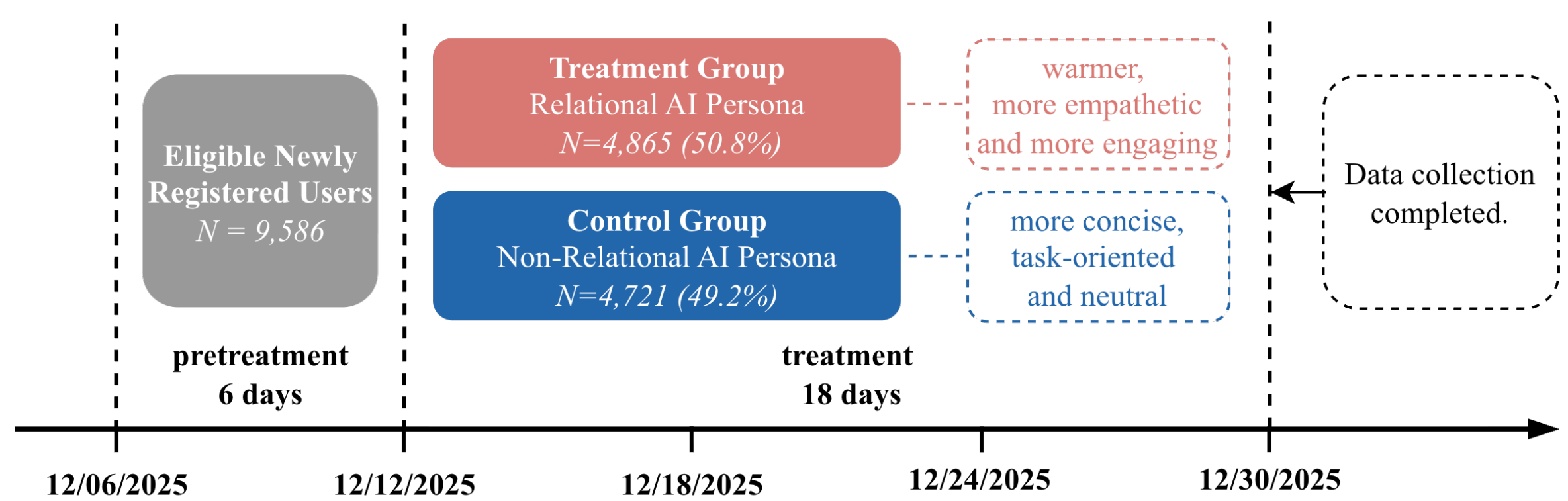

**Figure 2. Experiment Timeline**

### 3.5 Data Structure

Our raw data comprise three linked components: *user records*, *user–AI interaction records*, and *chat transcripts*. The user records contain account-level information, including treatment assignment, prior use of the AI service, and authentication provider. The interaction records capture timestamped activities of users and the chatbot, including file outputs. The chat transcripts contain the sender of each message and unstructured content data. We link the three components using unique user identifiers.

We construct the main analytical sample from the timestamped interaction records. The basic unit of service interaction is a *chat round*. Each user message marks the beginning of a new round, which includes all subsequent chatbot messages and file outputs generated before the user's next message. For each user, we order chat rounds chronologically and group them into *service sessions*. A new session begins when at least 30 minutes elapse between two consecutive chat rounds.[3] We then order each user's service sessions chronologically. The first service session is the earliest session observed during the treatment period, the second service session is the next observed session, and so forth. Collectively, all service sessions form a user's *conversation thread*. Because the platform supported only one conversation thread per user, the user ID and the conversation thread ID coincide. The resulting dataset therefore has a *nested hierarchical structure*, which is illustrated by Figure A1 in the Online Appendix: messages and file outputs are nested within chat rounds, chat rounds are nested within service sessions, and service sessions are nested within conversation threads, each corresponding to a unique user.

The final analytical sample contains 9,586 user-level conversation threads and 16,511 service sessions. Table 1 reports summary statistics for the main variables. On average, each conversation thread contains 1.722 service sessions, 10.771 chat rounds, and 2.682 file outputs. An average service session lasts 6.883 minutes and contains approximately 6.3 chat rounds. The small difference between chat rounds and total messages is because consecutive messages from the same sender are merged into a single message block.

## 4 Empirical Strategy

We examine how the relational AI persona affects service consumption along two dimensions: service interactions and service outputs. Section 4.1 describes the construction of the main dependent variables for each dimension. Section 4.2 introduces the key independent variables, while Section 4.3 defines the user intent dynamics measures used to examine the mechanisms underlying heterogeneous treatment effects. Sections 4.4 and 4.5 present the balance checks and empirical specifications, respectively.

### 4.1 Dependent Variables: Service Interactions and Outputs

We measure service consumption along two dimensions: *service interactions*, captured by *Sessions*, *Duration*, *Rounds*, and *User Intent Entropy*; and *service outputs*, captured by *Files* and *Goals*. We construct these outcomes at the user level for the main, first-session, and subsequent-session analyses, and at the user-period level for the dynamic analysis.

[3] Our main results are robust to alternative inactivity thresholds of 60 and 90 minutes. See Tables A6 and A7 for details.

**Service interactions.** *Sessions* counts the service sessions initiated by a user during the treatment period and captures how often the user engages with the chatbot. As described in Section 3, a new session begins when at least 30 minutes elapse between consecutive chat rounds. *Sessions* therefore captures the frequency of distinct service episodes. *Duration* sums the elapsed durations of these sessions, measured in minutes. Because its distribution is skewed, we use *ln(Duration)* in the regression analyses. *Rounds* counts user–AI chat rounds, each beginning with a user message and including the chatbot responses and file outputs generated before the next user message. Together, these measures capture the frequency, elapsed time, and volume of service interactions.

These measures do not describe the mix of intents pursued during an interaction. We therefore include *User Intent Entropy* to capture the diversity and relative prevalence of intents across chat rounds. To construct this measure, we classify user intent at the chat-round level using BERTopic, an unsupervised topic-modeling framework, followed by LLM-assisted labeling and manual review. This procedure distinguishes three substantive intent categories: *Task Execution* (performing a concrete task or producing an output), *Socialization* (engaging in social exchange, emotional expression, companionship, or role-playing), and *Knowledge Exploration* (seeking information, explanations, or ideas). Messages lacking substantive content are labeled *Non-Substantive*.[4] Appendix C provides full details of the classification procedure.

Let $\mathcal{S} = \{T, S, K\}$ denote the three substantive intent categories, i.e., *Task Execution*, *Socialization*, and *Knowledge Exploration*, respectively, and $n = |\mathcal{S}| = 3$. Let $p_{ia}$ denote the share of user *i*'s chat rounds assigned to intent category $a \in \mathcal{S}$. Following Shannon (1948), we define:

$$User\ Intent\ Entropy_i = -\frac{\sum_{a \in \mathcal{S}} p_{ia} \ln(p_{ia})}{\ln(n)}. \tag{1}$$

*User Intent Entropy* equals zero when all classified rounds express the same intent and reaches one when the rounds are equally distributed across the three categories. Higher values indicate a more diverse and even distribution of intents.

**Service outputs.** We measure service outputs using *Files* and *Goals*. The platform's output records associate generated files with goal identifiers (goal IDs), which group files linked to the same recorded goal. *Files* counts all files generated for a user, including images, videos, slide decks, and reports. Each file counts separately.

*Goals* counts the distinct goal IDs associated with these files. For example, five files sharing one goal ID contribute five to *Files* and one to *Goals*, whereas five files associated with five distinct goal IDs contribute five to both measures. Thus, *Files* captures the volume of file production, while *Goals* captures the number of distinct recorded goals associated with that production. Neither measure establishes successful completion of users' objectives.

[4] Non-substantive rounds contain isolated punctuation (e.g., "?") or random character strings (e.g., "asdfgh"). Including these rounds yields qualitatively similar results in the main and mechanism analyses (Table A2), suggesting that their exclusion does not drive our findings.

### 4.2 Independent Variables

Our primary independent variable is treatment assignment, denoted by $Treat_i$. It equals one if user $i$ was assigned to the relational AI persona and zero if the user was assigned to the non-relational persona. Control variables include user characteristics such as prior use of the chatbot and authentication method.

To examine variation by users' starting service intents, we construct *Entry Intent*, defined as the intent expressed in a session's opening user message. For the first-session heterogeneity analysis, we focus on sessions whose opening messages express a substantive intent: *Task Execution*, *Socialization*, or *Knowledge Exploration*. Whereas *Entry Intent* captures the session's starting intent, *User Intent Entropy* summarizes the distribution of substantive intents across its chat rounds.

For the first-session heterogeneity analysis, we interact treatment assignment with *Entry Intent* indicators, using *Task Execution* as the reference category. The opening user message precedes the first persona-conditioned AI response, providing the basis for using its intent as a pretreatment moderator.

As a robustness check, we construct *Majority Intent*, the most frequently observed intent among a session's substantive chat rounds. Whereas *Entry Intent* captures the starting service objective, *Majority Intent* captures the predominant intent expressed throughout a session. Section 6.2 compares the two classifications and reports the corresponding estimates.

### 4.3 Measures of User Intent Dynamics

*User Intent Entropy* summarizes the composition of intents but does not describe their sequence. To characterize how conversations unfold, we represent each user's first session as a sequence of intent states, following research on conversational state transitions (Cappella 1980, Liao et al. 2023). We construct two measures: *Intent Continuation Probability*, which captures the observed probability that consecutive rounds express the same substantive intent, and *Intent Transition Entropy*, which captures the variety of observed substantive intent-transition types. We use these measures as process evidence accompanying the heterogeneous effects.

Let $r = 1, \ldots, R_i$ denote the chat rounds in user $i$'s first session, where $R_i$ is the total number of rounds. Using the intent set $S$ defined in Section 4.1, let $s_{ir} \in S$ denote the intent of a substantive chat round. For any intent category $a, b \in S$, let $n_{iab}$ count eligible transitions between adjacent rounds from $s_{ir} = a$ to $s_{i,r+1} = b$.

Figure A2 illustrates this representation of transitions between adjacent intent states. The representation follows the structure used in Markov-based sequence analysis (Anderson and Goodman 1957) and analyses of conversational sequences (Cappella 1980). Because we summarize observed transitions rather than estimate a stochastic Markov process, our analysis does not require conversations to satisfy the memoryless property.

### 4.3.1 Intent Continuation Probability

*Intent Continuation Probability* measures the empirical probability that a pair of consecutive rounds, selected uniformly from a user's first session, expresses the same substantive intent. Let $N_{\mathrm{i}}^{\mathrm{self}} = \sum_{a\in\mathcal{S}} n_{iaa} = n_{iTT} + n_{iSS} + n_{iKK}$ denote the number of eligible same-intent transitions in user $i$'s first session. For $R_i > 1$, we define:

$$Intent\ Continuation\ Probability_i = \frac{N_i^{\mathrm{self}}}{R_i - 1}. \tag{2}$$

The denominator counts all adjacent-round pairs in the session. The measure ranges from zero to one: zero indicates no observed continuation within the same substantive intent category, while one indicates that every adjacent pair expresses the same substantive intent. Higher values therefore indicate greater intent continuation.

### 4.3.2 Intent Transition Entropy

*Intent Transition Entropy* measures the diversity and relative prevalence of observed transition types, including both same-intent transitions and changes in intent. For a session with at least one eligible transition, define:

$$q_{iab} = \frac{n_{iab}}{\sum_{a'\in\mathcal{S}} \sum_{b'\in\mathcal{S}} n_{ia'b'}}$$

as the share of eligible intent transitions. With three substantive intent categories, there are $n^2 = 9$ possible ordered transition types. Following Shannon (1948), we define:

$$Intent\ Transition\ Entropy_i = -\frac{\sum_{a\in\mathcal{S}} \sum_{b\in\mathcal{S}} q_{iab}\, ln(q_{iab})}{ln(n^2)}. \tag{3}$$

This measure equals zero when all valid transitions are of the same type and reaches one when all possible transition types occur in equal proportions. Higher values indicate greater transition variety.

## 4.4 Balance Check

As a balance check, we compare user characteristics and pretreatment service activity across treatment conditions. Table A1 reports these comparisons for prior use of the chatbot, authentication method, number of sessions, average session duration, chat rounds, user intent entropy, files created, and distinct goals served. None of these differences are statistically significant, with all reported p-values at least 0.261. These results are consistent with random assignment producing comparable groups on the observed baseline characteristics.

To assess the comparability of the groups used in the heterogeneity analysis, we also compare the distribution of first-session *Entry Intent* across treatment conditions. The proportions classified as *Task Execution*, *Socialization*, *Knowledge Exploration*, and *Non-Substantive* do not differ significantly between the relational and non-relational conditions, with p-values of 0.541, 0.201, 0.467, and 0.948, respectively. Thus, we find no statistically detectable differences in the observed composition of starting service objectives across conditions.

### 4.5 Empirical Specifications

We estimate the effect of relational persona assignment using the following baseline framework:

$$Y_i = \alpha_0 + \alpha_1 Treat_i + X_i'\gamma + \epsilon_i. \quad (4)$$

Here, $Y_i$ denotes the outcome variable for user $i$. For *ln(Duration)*, we estimate this linear specification using OLS. For *User Intent Entropy*, we use fractional logit regression with the same combination of independent variables. For the other count outcome variables, we use Poisson regression, in which the expected count equals the exponential of the same linear combination of independent variables. $Treat_i$ equals one if user $i$ was assigned to the relational AI persona and zero if the user was assigned to the non-relational AI persona. The coefficient of interest is $\alpha_1$, which captures the average treatment effect of assignment to the relational persona during the treatment period. The control vector, $X_i$, contains two pretreatment variables: an indicator for prior use of the chatbot and an indicator for authentication method. $\epsilon_i$ denotes the disturbance.

## 5 Empirical Results

We examine how relational persona design changes the amount, variety, and development of service consumption. Section 5.1 establishes the average effects on service interactions and outputs. Section 5.2 then examines how these effects vary with users' first-session entry intent. Section 5.3 then provides evidence on the mechanisms underlying these heterogeneous effects. Finally, Sections 5.4 and 5.5 extend the analysis beyond the first session by examining subsequent-session consumption and how the treatment effects evolve over time.

### 5.1 Average Treatment Effects of Relational AI Persona on Service Consumption

Table 2 reports the average treatment effects of assignment to the relational AI persona over the 18-day treatment period. We use Poisson regressions for the *number of sessions*, *total chat rounds*, *number of files created*, and *number of distinct goals served*; OLS for the log of *cumulative session duration*; and fractional logit regression for *user intent entropy*. For the count and log-duration outcomes, percentage changes are calculated using $100 \times (e^{\alpha_1} - 1)$. For the fractional logit model, percentage changes are based on average counterfactual predicted values under treatment and control.

**Service interactions.** Columns (1) and (2) show that the relational persona increases the number of sessions by 8.1% and cumulative session duration by 10.6% (both $p < 0.01$). Treated users therefore initiate more service sessions and spend more time in service over the treatment period. Columns (3) and (4) show that the relational persona increases both chat rounds and user intent entropy. The total number of chat rounds increases by 24.2% ($p < 0.01$), while user intent entropy increases by 5.8% ($p < 0.05$). Thus, treated users not only engage in more interactions but also distribute those interactions more broadly across intent categories.

**Service outputs.** Columns (5) and (6) show that the relational persona increases files created by 12.3% and distinct recorded goals by 12.1% (both $p < 0.01$). As defined in Section 4.1, these measures capture the volume of file production and the number of unique goals associated with that production. The relational persona therefore generates more files across more recorded goals.

Taken together, these results show that relational persona affects multiple dimensions of service consumption. Treated users initiate more sessions, spend more time in service, engage in more rounds across a more diverse mix of intents, and generate more outputs. These average effects, however, may conceal differences across users' starting intents. We next examine whether the increases in interactions and outputs vary with first-session *Entry Intent*.

### 5.2 Effects of Relational AI Persona on First-Session Service Consumption by User Entry Intent

To examine whether the effects of the relational persona vary with user entry intents, we focus on the first session and compare responses across *Task Execution*, *Socialization*, and *Knowledge Exploration*. We focus on the first session because its opening user message precedes the first persona-conditioned AI response. Table A1 reports no statistically significant differences in the distribution of first-session *Entry Intent* across treatment conditions.

We add *Entry Intent* indicators and their interactions with the treatment indicator, *Treat*, using *Task Execution* as the reference category. The treatment coefficient captures the effect for *Task Execution*, while the interaction coefficients measure differences relative to that category. Table 3 reports both the within-category treatment effects and the interaction estimates.

The analysis includes 9,092 users with first-session *Entry Intent*: 3,960 *Task Execution*, 3,248 *Socialization*, and 1,884 *Knowledge Exploration* users. We retain the same outcome framework as in Section 5.1. Because each user contributes one session, service interactions are measured by session duration, chat rounds, and user intent entropy. Service outputs are measured by files created and distinct recorded goals.

**Task Execution.** None of the estimated effects on session duration, chat rounds, user intent entropy, files created, or distinct goals served are statistically significant. We therefore find no statistically significant first-session treatment effect along these dimensions.

**Socialization.** The relational persona increases chat rounds by 16.0% ($p < 0.01$) and session duration by 10.8% ($p < 0.05$). Expected user intent entropy also increases by 9.2% ($p < 0.05$), indicating a more diverse distribution of interactions across intents. By contrast, neither the number of files created nor the number of distinct goals served changes significantly. Thus, greater interaction is accompanied by greater intent diversity, without a statistically significant increase in service outputs.

**Knowledge Exploration.** The relational persona increases total chat rounds by 15.9% ($p < 0.01$), session duration by 17.2%, files created by 18.7%, and distinct goals served by 18.9% (the latter three $p < 0.05$). User intent entropy does not change significantly. For these users, greater interaction accompanies more file production across more recorded goals without a detectable increase in intent diversity.[5]

These results reveal distinct patterns across the three *Entry Intent* categories. For *Task Execution* users, we find no significant treatment effects on service interactions or outputs. By contrast, both *Socialization* and *Knowledge Exploration* show similar increases in chat rounds and session durations. Similar increases in chat rounds, however, are associated with greater intent diversity for *Socialization* and greater output for *Knowledge Exploration*. These patterns motivate examining how intents unfold across successive rounds. We next investigate whether the additional interaction is accompanied by greater persistence within the same intent category or a broader distribution of intent transitions.

### 5.3 Heterogeneous First-Session User Intent Dynamics and Service Consumption

User intent entropy captures the overall variety of intents expressed in a session but does not describe their sequence. To characterize how user–AI interactions unfold, we examine the two measures introduced in Section 4.3. *Intent Continuation Probability* captures continuation within the same intent category, whereas *Intent Transition Entropy* captures the variety of observed intent-transition types, including same-intent transitions. These measures provide process evidence for interpreting the heterogeneous consumption effects. Figure A2 provides a descriptive view of transitions between adjacent chat rounds.

Table 4 reports estimates using the same first-session *Entry Intent* categories as Table 3. For *Task Execution*, neither *Intent Continuation Probability* nor *Intent Transition Entropy* changes significantly. For *Socialization*, *Intent Transition Entropy* increases by 11.8% ($p < 0.05$), while *Intent Continuation Probability* does not change significantly. Together with the increase in user intent entropy reported in Table 3, this pattern is consistent with a broader distribution of intents and more varied transitions.

*Knowledge Exploration* shows a different pattern. *Intent Continuation Probability* increases by 12.8% ($p < 0.1$), while neither *Intent Transition Entropy* nor *User Intent Entropy* changes significantly. This provides tentative evidence of greater continuation within the same intent category, without a significant increase in intent diversity.

[5] We also examine repeated file creation for the same goal. The estimated treatment effects are statistically insignificant for all three *Entry Intent* categories (see Table A3).

These results help characterize the consumption patterns documented in Section 5.2. For *Socialization*, greater interaction accompanies broader intent composition and more varied transitions, without a significant increase in service outputs. For *Knowledge Exploration*, greater interactions accompany an increase in intent continuation, as well as elevated outputs.

### 5.4 Effects of the Relational AI Persona on Subsequent-Session Service Consumption

We next examine how the effects of the relational persona extend beyond the first session. Table 5 classifies each subsequent session by its *Entry Intent* and aggregates outcomes by intent at the user level. All regressions retain the full sample of 9,586 users, with zero counts when a user has no subsequent session of a given intent. The key objective is to estimate the likelihood of initiating subsequent sessions with different entry intents, the number of sessions, durations, chat rounds, intent entropy, and outputs.

For *Task Execution* and *Socialization*, the relational persona increases both the probability and number of subsequent sessions. Session counts increase by 12.2% and 49.4%, respectively (both $(p < 0.01)$), while the probability of having at least one subsequent session increases by 1.5 and 1.2 percentage points (both $(p < 0.05)$). The session count estimate is larger for *Socialization*. Both categories also show significant increases in cumulative session duration, chat rounds, user intent entropy, files created, and distinct recorded goals.

For *Knowledge Exploration,* the effects on subsequent-session probability, session counts, and cumulative duration are statistically insignificant. Nevertheless, aggregate chat rounds, files created, and distinct recorded goals increase significantly. User intent entropy does not change significantly. This pattern resembles the first-session results in that greater interaction accompanies greater output without an increase in intent diversity.

The subsequent-session results therefore extend the first-session findings: output gains appear across all three substantive intent categories, while significant increases in session probability and counts occur for *Task Execution* and *Socialization*. For *Knowledge Exploration*, aggregate interactions and outputs increase without a statistically significant increase in session counts.

### 5.5 Dynamic Effects of the Relational AI Persona across Six-Day Intervals

Finally, Table 6 examines how treatment effects vary over the 18-day treatment period. We divide the period into three six-day intervals, denoted *Periods 1, 2, and 3*, with *Period 1* as the reference. The regressions include period fixed effects and user-clustered standard errors, yielding 28,758 user-period observations.

**Service interactions.** The relative effects on session counts are larger after the first period. Session counts increase by 3.4% in *Period 1*, 25.4% in *Period 2*, and 30.4% in *Period 3*. Both later-period effects exceed the *Period 1* effect at the 10% significance level. Expected user intent entropy increases by 4.6%, 56.3%, and 51.9%, respectively, with positive and statistically significant treatment-by-period interactions for both later periods.

Chat rounds increase by 16.2%, 56.0%, and 47.9% across the three periods. Cumulative session duration follows a different pattern: its estimated increase declines from 9.3% in *Period 1* to 6.4% in *Period 2* and 2.6% in *Period 3*, with the final-period effect significantly smaller than the initial effect. Thus, the duration effect remains positive, even as its magnitude declines.

**Service outputs.** The point estimates imply increases in files created of 9.2%, 19.1%, and 21.4% across the three periods, and increases in distinct recorded goals of 8.3%, 19.4%, and 25.7%. Although the later-period point estimates are larger, neither later-period output effect is statistically significant, and the differences across periods are also statistically insignificant.

Together, later in the treatment period, the relational persona has larger effects on session counts and intent diversity, but a smaller effect on cumulative duration. Its effects on service outputs do not increase significantly over time.

## 6 Robustness Check

### 6.1 Robustness to an Alternative Empirical Model

To examine whether our findings depend on the main empirical specifications, we re-estimate the average treatment effects using OLS for all outcome variables. The analysis uses the full sample and the same controls as the main specification. The service output measures are winsorized at the 99th percentile.

Table A4 reports consistent results. The relational persona increases the number of sessions by 0.134, cumulative session duration by 3.674 minutes, total chat rounds by 2.326, and user intent entropy by 0.015. It also increases files created by 0.222 and distinct goals served by 0.186. All estimates are statistically significant at the 5% level or lower.

Overall, the OLS estimates are consistent in direction and statistical significance with the main results, showing that the effects on service interactions and outputs are robust to alternative model specifications.

### 6.2 Robustness to an Alternative Definition of Session Intent

Our main analysis classifies each session by *Entry Intent*, defined as the user intent expressed in the first chat round. To assess whether the results depend on this definition, we construct an alternative measure, *Majority Intent*, defined as the most frequently observed substantive intent across all classified chat rounds within a session. Among the 6,925 sessions in the subsequent-session analysis, both measures yield a substantive intent classification for 5,778 sessions and agree for 5,419, corresponding to an agreement rate of 93.79%. Table A5 re-estimates the subsequent-session analysis using *Majority Intent*. The results closely mirror those based on *Entry Intent*.

For service interactions, the relational persona increases the probability of subsequent *Task Execution* and *Socialization* sessions by 1.5 percentage points in both categories, and increases their session counts by 11.9% and 55.9%, respectively. In contrast, neither the probability nor the number of subsequent *Knowledge Exploration* sessions increases significantly. Chat rounds increase across all three substantive intent categories. User intent entropy increases by 17.6% for *Task Execution* ($p < 0.10$) and 25.7% for *Socialization* ($p < 0.05$), while the effect for *Knowledge Exploration* remains statistically insignificant, consistent with the main analysis.

The results for service outputs are similarly robust. Files created and distinct goals served also increase significantly across all three categories. Overall, the alternative intent definition preserves the main patterns in subsequent service interactions and service outputs.

Because *Majority Intent* incorporates chat rounds observed after the session begins, it may itself be influenced by the assigned persona. We therefore use it only as an alternative classification for subsequent-session consumption rather than as a moderator in the first-session heterogeneity analysis. The high agreement between the two intent measures and the consistency of the estimated effects indicate that our findings are not driven by defining session intent solely from the opening chat round.

### 6.3 Robustness to Alternative Session Inactivity Thresholds

Our main analysis defines a new session after 30 minutes of inactivity. To examine whether the results depend on this choice, we reconstruct all session-based measures using 60- and 90-minute inactivity thresholds. Tables A6 and A7 report the results.

The average treatment effects remain stable. The relational persona increases the number of sessions by 7.3% and 7.1% under the 60- and 90-minute thresholds, respectively, and cumulative session duration by 11.4% and 10.7% (all $p < 0.01$). Outcomes that do not depend on session boundaries, including total chat rounds, user intent entropy, and service outputs, remain unchanged.

The first-session heterogeneity results also closely resemble the main findings. For *Socialization* users, chat rounds increase by 18.2% and 16.9%, cumulative duration by 11.1% and 10.4%, and user intent entropy by 8.5% and 8.3% under the two alternative thresholds. Service output effects remain limited. For *Knowledge Exploration* users, chat rounds increase by 18.0% and 16.1%, duration by 19.2% and 17.5%, and files and goals by approximately 20% under both thresholds, while user intent entropy remains insignificant. The intent-dynamics results are also preserved: *Socialization* users continue to exhibit greater intent transition entropy, while *Knowledge Exploration* users exhibit greater intent continuation probability. Effects for *Task Execution* remain limited overall.

The subsequent-session results show the same qualitative pattern. The relational persona increases subsequent session frequency for *Task Execution* and especially *Socialization*, but not significantly for *Knowledge Exploration*. At the same time, chat rounds increase across all three substantive intent categories, user intent entropy increases for *Task Execution* and *Socialization* but not *Knowledge Exploration*, and service outputs increase across all three categories. Overall, the main conclusions are robust to alternative inactivity thresholds used to define service sessions.

### 6.4 Human Validation of Classification of Session Intent

To assess the reliability of our session-intent measure, we conduct a human validation exercise using a stratified random sample of 1,600 first sessions. We sample 200 observations from each treatment-by-intent stratum to ensure sufficient representation of all intent categories. Two trained research assistants independently classify each session using the same intent definitions while remaining blind to treatment assignment and LLM-generated labels. Disagreements are resolved through discussion or adjudication by a third reviewer.

The algorithm-generated classification agrees with the adjudicated human classification in 83.9% of cases, corresponding to a 16.1% error rate. Inter-annotator agreement is also high, with a Cohen's Kappa of 0.842 (Cohen 1960). Category-specific precision ranges from 72.3% to 90.5%, while recall ranges from 77.3% to 91.6%. Importantly, classification error is nearly identical across treatment conditions: 16.3% under the relational persona and 16.0% under the non-relational persona, with no significant difference ($p = 0.892$).

Most disagreements occur for brief, ambiguous, or multi-intent opening messages, suggesting that errors arise primarily near the conceptual boundaries between intent categories rather than from a systematic tendency to favor a particular category or treatment condition. Overall, the validation exercise indicates that the session-intent measure has high reliability and that classification error is unlikely to systematically bias the estimated heterogeneous treatment effects.

## 7 Managerial Implications and Conclusions

LLMs are changing digital service operations from a process in which customers follow predefined steps to one that develops through open-ended conversation. This shift makes persona an operational design choice. By shaping how the AI communicates, persona design can influence how often and how long users access the service, how much and how broadly they interact, and what service outputs are generated.

Using a randomized field experiment involving 9,586 users on an LLM platform, we compare a relational persona with a non-relational persona while holding the underlying LLM and service capabilities constant. The relational persona increases service consumption across two dimensions: service interactions (service sessions, +8.1%; cumulative duration, +10.6%; chat rounds, +24.2%; user intent entropy, +5.8%) and service outputs (files, +12.3%; distinct goals, +12.1%). These findings show that persona design affects actual service consumption rather than merely users' perceptions of the interface.

These average effects, however, conceal important differences across users' first-session entry intents. The relational persona produces no statistically significant first-session effects for *Task Execution* users. For *Socialization* users, it increases interactions but not service outputs. For *Knowledge Exploration* users, it produces a nearly identical increase in chat rounds but also increases service outputs without increasing user intent entropy. To characterize the mechanism behind these heterogeneous patterns, we represent user intent dynamics as a transition process of intent states. For *Socialization* users, the relational persona increases intent transition entropy by 11.8%, indicating that the interaction unfolds across a wider range of intent transitions. For *Knowledge Exploration* users, it instead increases intent continuation probability by 12.8%, indicating more persistent engagement with the same service intent. Thus, similar increases in interaction volume can follow different conversational paths and be accompanied by different service output responses.

These heterogeneous patterns largely persist in subsequent use. However, subsequent session frequency increases most for *Socialization*, increases more modestly for *Task Execution*, but does not increase significantly for *Knowledge Exploration*, despite substantial increases in total chat rounds and service outputs. Our dynamic effects analysis further shows that, over time, the relational persona has larger effects on session counts and intent diversity, but a smaller effect on cumulative duration.

For managers, persona should be treated as part of the service operating system rather than merely as a presentation feature. Because its effects vary across service objectives, applying the same persona universally may be suboptimal. A relational persona can substantially expand *Socialization* use, generate greater interaction and output in *Knowledge Exploration*, and have more limited immediate effects on well-defined *Task Execution*. Furthermore, because greater interaction does not uniformly translate into greater output, firms should evaluate service interactions and service outputs separately, while considering the costs of computation.

Our study has several limitations that also point to directions for future research. We examine one LLM platform, one composite relational persona intervention, and newly registered users over an 18-day treatment period. Future work could test whether these effects generalize across models, service settings, user populations, and longer time horizons, as well as isolate specific persona dimensions such as warmth, empathy, and anthropomorphism. More broadly, research could examine how open-ended LLM services should structure evolving user objectives and whether personas can be dynamically adapted to different service intents. As conversational AI becomes a primary interface for digital service delivery, understanding how to design not only what an LLM can do, but also how it interacts, adapts, and guides service consumption, will become a central question in LLM service operations.

## References

**Agnes AI. (2025).** 150,000 daily users since launch in July: Agnes AI surpasses key milestone. *GlobeNewswire*. Retrieved from: https://www.globenewswire.com/news-release/2025/11/04/3179935/0/en/150-000-Daily-Users-Since-Launch-in-July-Agnes-AI-Surpasses-Key-Milestone.html

**Anderson, T. W., & Goodman, L. A. (1957).** Statistical inference about Markov chains. *The Annals of Mathematical Statistics*, 28(1), 89–110.

**Balakrishnan, M., Ferreira, K. J., & Tong, J. (2025).** Human-algorithm collaboration with private information:

Naïve advice-weighting behavior and mitigation. *Management Science*, *72*(1), 265-284.

**Bastani, H., Bastani, O., Sungu, A., Ge, H., Kabakcı, Ö., & Mariman, R. (2025).** Generative AI without guardrails can harm learning: Evidence from high school mathematics. *Proceedings of the National Academy of Sciences*, 122(26), e2422633122.

**Bellos, I., & Kavadias, S. (2020).** Service design for a holistic customer experience: A process framework. *Management Science*, 67(3), 1718–1736.

**Bergner, A. S., Hildebrand, C., Häubl, G., Inman, J. J., Lutz, R. J., & Kyung, E. J. (2023).** Machine talk: How verbal embodiment in conversational AI shapes consumer-brand relationships. *Journal of Consumer Research*, 50(4), 742–764.

**Blut, M., Wang, C., Wünderlich, N. V., & Brock, C. (2021).** Understanding anthropomorphism in service provision: A meta-analysis of physical robots, chatbots, and other AI. *Journal of the Academy of Marketing Science*, 49, 632–658.

**Brynjolfsson, E., Li, D., & Raymond, L. (2025).** Generative AI at work. *The Quarterly Journal of Economics*, 140(2), 889–942.

**Campbell, D., & Frei, F. X. (2009).** Cost structure, customer profitability, and retention implications of self-service distribution channels: Evidence from customer behavior in an online banking channel. *Management Science*, 56(1), 4–24.

**Cappella, J. N. (1980).** Talk and silence sequences in informal conversations II. *Human Communication Research*, 6(2), 130–145.

**Cohen, J. (1960).** A coefficient of agreement for nominal scales. *Educational and Psychological Measurement*, 20(1), 37–46.

**Dixon, M. J., & Verma, R. (2013).** Sequence effects in service bundles: Implications for service design and scheduling. *Journal of Operations Management*, 31(3), 138–152.

**Dixon, M. J., Victorino, L., Kwortnik, R. J., & Verma, R. (2017).** Surprise, anticipation, and sequence effects in the design of experiential services. *Production and Operations Management*, 26(5), 945–960.

**Fang, Z., Chang, Y., Luo, X., Wu, Q., & Aspara, J. (2026).** Artificial intelligence, emotional labor, and service operations. *Manufacturing & Service Operations Management*.

**Go, E., & Sundar, S. S. (2019).** Humanizing chatbots: The effects of visual, identity and conversational cues on humanness perceptions. *Computers in Human Behavior*, 97, 304–316.

**Grootendorst, M. (2022).** BERTopic: Neural topic modeling with a class-based TF-IDF procedure. *arXiv* preprint arXiv:2203.05794.

**Heinz, M. V., Mackin, D. M., Trudeau, B. M., Bhattacharya, S., Wang, Y., Banta, H. A., … & Jacobson, N. C. (2025).** Randomized trial of a generative AI chatbot for mental health treatment. *NEJM AI*, 2(4),

Aloa2400802.

**Hugging Face. (2026).** Agnes-SeaLLM-8B. Retrieved from: https://huggingface.co/Agnes-AI/Agnes-SeaLLM-8b

**Ibrahim, L., Hafner, F. S., & Rocher, L. (2026).** Training language models to be warm can reduce accuracy and increase sycophancy. *Nature*, 652, 1159–1165.

**Kwon, D., Shrestha, K., Han, B., Lee, E. H., & Lucas, G. (2025).** Evaluating behavioral alignment in conflict dialogue: A multi-dimensional comparison of LLM agents and humans. *Proceedings of the 2025 Conference on Empirical Methods in Natural Language Processing*, 16366–16380.

**Lemon, K. N., & Verhoef, P. C. (2016).** Understanding customer experience throughout the customer journey. *Journal of Marketing*, 80(6), 69–96.

**Liao, W., Oh, Y. J., Zhang, J., & Feng, B. (2023).** Conversational dynamics of joint attention and shared emotion predict outcomes in interpersonal influence situations: An interaction ritual perspective. *Journal of Communication*, 73(4), 342–355.

**Luo, X., Tong, S., Fang, Z., & Qu, Z. (2019).** Frontiers: Machines vs. humans: The impact of artificial intelligence chatbot disclosure on customer purchases. *Marketing Science*, 38(6), 937–947.

**Meuter, M. L., Ostrom, A. L., Roundtree, R. I., & Bitner, M. J. (2000).** Self-service technologies: Understanding customer satisfaction with technology-based service encounters. *Journal of Marketing*, 64(3), 50–64.

**Ni, X., Wang, Y., Feng, T., Lu, L. X., Wang, Y., & Zhou, C. (2026).** Generative AI in action: Field experimental evidence from Alibaba's customer service operations. Available at *SSRN*: 5012601.

**Noy, S., & Zhang, W. (2023).** Experimental evidence on the productivity effects of generative artificial intelligence. *Science*, *381*(6654), 187–192.

**Sapiens AI. (2025).** SapiensAI launches Agnes – Singapore's homegrown answer to DeepSeek. *PR Newswire*. Retrieved from: https://www.prnewswire.com/apac/news-releases/sapiensai-launches-agnes--singapores-homegrown-answer-to-deepseek-302443885.html

**Schanke, S., Burtch, G., & Ray, G. (2021).** Estimating the impact of "humanizing" customer service chatbots. *Information Systems Research*, 32(3), 736–751.

**Shannon, C. E. (1948).** A mathematical theory of communication. *Bell System Technical Journal*, 27(3), 379–423; 27(4), 623–656.

**Tu, Q., Fan, S., Tian, Z., Shen, T., Shang, S., Gao, X., & Yan, R. (2024).** CharacterEval: A Chinese benchmark for role-playing conversational agent evaluation. *Proceedings of the 62nd Annual Meeting of the Association for Computational Linguistics (Volume 1: Long Papers)*, 11836–11850.

**Wang, L., Yang, N., Huang, X., Yang, L., Majumder, R., & Wei, F. (2024).** Multilingual E5 text embeddings:

A technical report. *arXiv* preprint arXiv:2402.05672.

**Wang, Y., Zhu, C., Feng, T., Lu, L. X., & Jia, B. (2026).** Agentic AI and human-in-the-loop interventions: Field experimental evidence from Alibaba's customer service operations. Available at *SSRN*: 6745878.

**Xu, Y., Dai, H., & Yan, W. (2024).** Identity disclosure and anthropomorphism in voice chatbot design: A field experiment. *Management Science*, *72*(1), 223–241.

**Xue, M., Hitt, L. M., & Chen, P.-Y. (2011).** Determinants and outcomes of internet banking adoption. *Management Science*, 57(2), 291–307.

**Xue, M., Hitt, L. M., & Harker, P. T. (2007).** Customer efficiency, channel usage, and firm performance in retail banking. *Manufacturing & Service Operations Management*, 9(4), 535–558.

**Zhang, P., Cui, R., & Zhang, D. J. (2026).** The impact of AI search on the online content ecosystem: Evidence from Google and Reddit. *arXiv* preprint arXiv:2605.16428.

## Table 1. Summary Statistics

| Variable | Definition | N | Mean | SD |
|---|---|---|---|---|
| **Panel A: User Characteristic Variables** | | | | |
| Treat | 1 = Relational AI; 0 = Non-Relational AI | 9,586 | 0.508 | 0.500 |
| Prior AI Use | 1 for users who previously used the AI tool; 0 otherwise | 9,586 | 0.111 | 0.314 |
| If Login Type = Google | 1 when the authentication provider is Google; 0 otherwise | 9586 | 0.879 | 0.326 |
| **Panel B: Outcome Variables Across All Service Sessions** | | | | |
| Total User Messages | # of user-sent messages | 9,586 | 11.176 | 42.624 |
| Total Characters in User Messages | # of characters in user-sent messages | 9,586 | 1095.2 | 6009.7 |
| Total AI Messages | # of AI-sent messages | 9,586 | 18.976 | 63.097 |
| Total Characters in AI Messages | # of characters in AI-sent messages | 9,586 | 4058.3 | 14196.9 |
| Number of Sessions | # of service sessions in a user's conversation thread | 9,586 | 1.722 | 2.836 |
| Total Session Duration (minutes) | Total session duration, measured in minutes | 9,586 | 17.569 | 71.533 |
| Average Session Duration (minutes) | Average duration per session, measured in minutes | 9,586 | 6.883 | 11.566 |
| Chat Rounds | # of user–AI message exchanges in a chat | 9,586 | 10.771 | 41.448 |
| User Intent Entropy | Variety of a user's intents across chat rounds | 9,586 | 0.269 | 0.317 |
| Files Created | # of files created by AI, i.e., Image, Video, Slide Deck, and Report | 9,586 | 2.682 | 9.750 |
| Goals Served | # of distinct platform-assigned goal IDs observed for a user | 9,586 | 2.345 | 8.534 |
| **Panel C: Outcome Variables in the First Service Session** | | | | |
| *Entry Intent = Task Execution (N=3,960)* | | | | |
| Duration (minutes) | Interaction duration among users whose intent is task execution | 3,960 | 10.600 | 16.779 |
| Chat Rounds | # of chat rounds among users whose intent is task execution | 3,960 | 6.231 | 8.564 |
| User Intent Entropy | Variety of intents among users whose intent is task execution | 3,960 | 0.266 | 0.310 |
| Files Created | # of file outputs among users whose intent is task execution | 3,960 | 2.223 | 3.058 |
| Goals Served | # of goals served among users whose intent is task execution | 3,960 | 1.990 | 2.753 |
| *Entry Intent = Socialization (N=3,248)* | | | | |
| Duration (minutes) | Interaction duration among users whose intent is socialization | 3,248 | 6.786 | 16.935 |
| Chat Rounds | # of chat rounds among users whose intent is socialization | 3,248 | 6.001 | 12.992 |
| User Intent Entropy | Variety of intents among users whose intent is socialization | 3,248 | 0.255 | 0.317 |
| Files Created | # of file outputs among users whose intent is socialization | 3,248 | 0.665 | 1.896 |
| Goals Served | # of goals served among users whose intent is socialization | 3,248 | 0.559 | 1.488 |
| *Entry Intent = Knowledge Exploration (N=1,884)* | | | | |
| Duration (minutes) | Interaction duration among users whose intent is knowledge exploration | 1,884 | 6.023 | 13.958 |
| Chat Rounds | # of chat rounds among users whose intent is knowledge exploration | 1,884 | 4.456 | 6.829 |
| User Intent Entropy | Variety of intents among users whose intent is knowledge exploration | 1,884 | 0.211 | 0.311 |
| Files Created | # of file outputs among users whose intent is knowledge exploration | 1,884 | 0.470 | 1.278 |
| Goals Served | # of goals served among users whose intent is knowledge exploration | 1,884 | 0.400 | 1.039 |

***Notes.*** The unit of observation is the user. Entry intent is classified from the first valid user message before the first persona-conditioned AI response. First-session entry intent is identified for 9,092 of the 9,586 users, with the remaining 494 users excluded due to non-substantive entry messages.

## Table 2. Average Treatment Effects of the Relational AI Persona on Service Consumption

| | *Service Interactions* | | | | *Service Outputs* | |
|---|---|---|---|---|---|---|
| | Number of Sessions | ln(Cumulative Session Duration) | Total Chat Rounds | User Intent Entropy | Files Created | Goals Served |
| | (1) | (2) | (3) | (4) | (5) | (6) |
| Treat | 0.078*** | 0.101*** | 0.217*** | 0.077** | 0.116*** | 0.114*** |
| | (0.016) | (0.033) | (0.006) | (0.033) | (0.013) | (0.013) |
| *Δ Percentage* | 8.1%*** | 10.6%*** | 24.2%*** | 5.8%** | 12.3%*** | 12.1%*** |
| Controls | Y | Y | Y | Y | Y | Y |
| Observations | 9,586 | 9,586 | 9,586 | 9,586 | 9,586 | 9,586 |
| (Pseudo) $R^2$ | 0.020 | 0.015 | 0.042 | 0.001 | 0.018 | 0.019 |

***Notes.*** The omitted treatment condition is the non-relational AI persona. Count outcomes are estimated using Poisson regression. Log-transformed continuous outcomes are estimated using ordinary least squares (OLS) regression. Bounded index outcomes are estimated using fractional logit regression. Standard errors are in parentheses. *p<0.1; **p<0.05; ***p<0.01.

**Table 3. Effects of the Relational AI Persona on First-Session Service Consumption by First-Session Entry Intent**

| | *Service Interactions* | | | *Service Outputs* | |
|---|---|---|---|---|---|
| | ln(Session Duration) | Chat Rounds | User Intent Entropy | Files Created | Goals Served |
| | (1) | (2) | (3) | (4) | (5) |
| Treat | 0.063 | 0.007 | 0.049 | 0.025 | 0.009 |
| | (0.044) | (0.013) | (0.052) | (0.021) | (0.023) |
| Treat × Socialization | 0.039 | 0.141*** | 0.069 | 0.003 | 0.067 |
| | (0.066) | (0.019) | (0.078) | (0.048) | (0.052) |
| Treat × Knowledge Exploration | 0.096 | 0.141*** | -0.030 | 0.147** | 0.163** |
| | (0.078) | (0.025) | (0.097) | (0.071) | (0.077) |
| Treatment Effect: Entry Intent = Task Execution | 0.063 | 0.007 | 0.049 | 0.025 | 0.009 |
| Treatment Effect: Entry Intent = Socialization | 0.103** | 0.149*** | 0.118** | 0.028 | 0.077 |
| Treatment Effect: Entry Intent = Knowledge Exploration | 0.159** | 0.148*** | 0.019 | 0.172** | 0.173** |
| Δ Percentage if Entry Intent = Task Execution | 6.5% | 0.7% | 3.7% | 2.5% | 0.9% |
| Δ Percentage if Entry Intent = Socialization | 10.8%** | 16.0%*** | 9.2%** | 2.8% | 8.0% |
| Δ Percentage if Entry Intent = Knowledge Exploration | 17.2%** | 15.9%*** | 1.5% | 18.7%** | 18.9%** |
| Controls | Y | Y | Y | Y | Y |
| Observations | 9,092 | 9,092 | 9,092 | 9,092 | 9,092 |
| (Pseudo) $R^2$ | 0.044 | 0.010 | 0.002 | 0.122 | 0.129 |

***Notes.*** The reference category for session intent is *Task Execution* . First-session intent is identified for 9,092 of the 9,586 users, with the remaining 494 users excluded due to unclear intent. Session entry intent classification is balanced across conditions. Count outcomes are estimated using Poisson regression. Log-transformed continuous outcomes are estimated using OLS regression. Bounded index outcomes are estimated using fractional logit regression. Standard errors are in parentheses. *p<0.1; **p<0.05; ***p<0.01.

**Table 4. Effects of the Relational AI Persona on First-Session User Intent Dynamics by First-Session Entry Intent**

| | *First-Session User Intent Dynamics* | |
|---|---|---|
| | Intent Continuation Probability | Intent Transition Entropy |
| | (1) | (2) |
| Treat | -0.001 | 0.022 |
| | (0.049) | (0.053) |
| Treat × Socialization | -0.002 | 0.110 |
| | (0.076) | (0.081) |
| Treat × Knowledge Exploration | 0.157 | 0.010 |
| | (0.097) | (0.107) |
| Treatment Effect: Entry Intent = Task Execution | -0.001 | 0.022 |
| Treatment Effect: Entry Intent = Socialization | -0.003 | 0.132** |
| Treatment Effect: Entry Intent = Knowledge Exploration | 0.156* | 0.032 |
| Δ Percentage if Entry Intent = Task Execution | -0.1% | 1.8% |
| Δ Percentage if Entry Intent = Socialization | -0.2% | 11.8%** |
| Δ Percentage if Entry Intent = Knowledge Exploration | 12.8%* | 2.8% |
| Controls | Y | Y |
| Observations | 9,092 | 9,092 |
| Pseudo $R^2$ | 0.016 | 0.006 |

***Notes.*** The reference category for session intent is *Task Execution* . First-session intent is identified for 9,092 of the 9,586 users, with the remaining 494 users excluded due to unclear intent. Session entry intent classification is balanced across conditions. All results are estimated using fractional logit regression. Standard errors are in parentheses. *p<0.1; **p<0.05; ***p<0.01.

## Table 5. Effects of the Relational AI Persona on Subsequent-Session Service Consumption

**Panel A: Any Subsequent Session**

| | Task Execution | Socialization | Knowledge Exploration | Non-Substantive |
|---|---|---|---|---|
| | (1) | (2) | (3) | (4) |
| Treat | 0.129** | 0.201** | 0.046 | 0.249*** |
| | (0.060) | (0.086) | (0.111) | (0.084) |
| *Change in Odds (%)* | 13.8%** | 22.3%** | 4.8% | 28.3%*** |
| *Average Marginal Effect (pp)* | +1.5 pp** | +1.2 pp** | +0.2 pp | +1.5 pp*** |
| Controls | Y | Y | Y | Y |
| Observations | 9,586 | 9,586 | 9,586 | 9,586 |
| Pseudo $R^2$ | 0.013 | 0.018 | 0.027 | 0.031 |

**Panel B: Number of Subsequent Sessions**

| | Task Execution | Socialization | Knowledge Exploration | Non-Substantive |
|---|---|---|---|---|
| | (1) | (2) | (3) | (4) |
| Treat | 0.115*** | 0.402*** | 0.089 | 0.247*** |
| | (0.032) | (0.058) | (0.080) | (0.060) |
| *Δ Percentage* | 12.2%*** | 49.4%*** | 9.3% | 28.0%*** |
| Controls | Y | Y | Y | Y |
| Observations | 9,586 | 9,586 | 9,586 | 9,586 |
| Pseudo $R^2$ | 0.024 | 0.054 | 0.035 | 0.049 |

**Panel C: ln(Cumulative Session Duration)**

| | Task Execution | Socialization | Knowledge Exploration | Non-Substantive |
|---|---|---|---|---|
| | (1) | (2) | (3) | (4) |
| Treat | 0.045** | 0.030** | 0.008 | 0.036*** |
| | (0.021) | (0.013) | (0.009) | (0.013) |
| *Δ Percentage* | 4.6%** | 3.0%** | 0.8% | 3.7%*** |
| Controls | Y | Y | Y | Y |
| Observations | 9,586 | 9,586 | 9,586 | 9,586 |
| $R^2$ | 0.009 | 0.008 | 0.009 | 0.013 |

**Panel D: Subsequent Total Chat Rounds and User Intent Entropy**

| | Task Execution | | Socialization | | Knowledge Exploration | | Non-Substantive | |
|---|---|---|---|---|---|---|---|---|
| | Total Chat Rounds | User Intent Entropy | Total Chat Rounds | User Intent Entropy | Total Chat Rounds | User Intent Entropy | Total Chat Rounds | User Intent Entropy |
| | (1) | (2) | (3) | (4) | (5) | (6) | (7) | (8) |
| Treat | 0.223*** | 0.238** | 0.499*** | 0.327*** | 0.475*** | 0.107 | 0.436*** | 0.085 |
| | (0.014) | (0.106) | (0.018) | (0.114) | (0.030) | (0.156) | (0.022) | (0.142) |
| *Δ Percentage* | 25.0%*** | 26.1%** | 64.7%*** | 37.6%*** | 60.8%*** | 11.1% | 54.7%*** | 8.7% |
| Controls | Y | Y | Y | Y | Y | Y | Y | Y |
| Observations | 9,586 | 9,586 | 9,586 | 9,586 | 9,586 | 9,586 | 9,586 | 9,586 |
| Pseudo $R^2$ | 0.036 | 0.014 | 0.081 | 0.015 | 0.071 | 0.012 | 0.066 | 0.027 |

**Panel E: Subsequent Service Outputs**

| | Task Execution | | Socialization | | Knowledge Exploration | | Non-Substantive | |
|---|---|---|---|---|---|---|---|---|
| | Files Created | Goals Served | Files Created | Goals Served | Files Created | Goals Served | Files Created | Goals Served |
| | (1) | (2) | (3) | (4) | (5) | (6) | (7) | (8) |
| Treat | 0.139*** | 0.154*** | 0.515*** | 0.483*** | 1.221*** | 1.224*** | 0.294*** | 0.204** |
| | (0.018) | (0.020) | (0.079) | (0.085) | (0.124) | (0.134) | (0.076) | (0.081) |
| *Δ Percentage* | 15.0%*** | 16.6%*** | 67.3%*** | 62.1%*** | 239.0%*** | 240.1%*** | 34.2%*** | 22.7%** |
| Controls | Y | Y | Y | Y | Y | Y | Y | Y |
| Observations | 9,586 | 9,586 | 9,586 | 9,586 | 9,586 | 9,586 | 9,586 | 9,586 |
| Pseudo $R^2$ | 0.033 | 0.033 | 0.034 | 0.034 | 0.099 | 0.095 | 0.030 | 0.033 |

***Notes.*** Count outcomes are estimated using Poisson regression. Log-transformed continuous outcomes are estimated using OLS regression. Bounded index outcomes are estimated using fractional logit regression. Binary outcomes are estimated using logit regression. Standard errors are given in parentheses. *p<0.1; **p<0.05; ***p<0.01.

## Table 6. Dynamic Effects of the Relational AI Persona across Six-Day Intervals

| | *Service Interactions* | | | | *Service Outputs* | |
|---|---|---|---|---|---|---|
| | Number of Sessions | ln(Cumulative Session Duration) | Total Chat Rounds | User Intent Entropy | Files Created | Goals Served |
| | (1) | (2) | (3) | (4) | (5) | (6) |
| Treat | 0.033* | 0.089*** | 0.150*** | 0.062* | 0.088* | 0.080 |
| | (0.019) | (0.031) | (0.057) | (0.033) | (0.051) | (0.050) |
| Treat × Days 7–12 | 0.193* | -0.027 | 0.295** | 0.394*** | 0.087 | 0.098 |
| | (0.103) | (0.031) | (0.139) | (0.118) | (0.156) | (0.157) |
| Treat × Days 13–18 | 0.232* | -0.063** | 0.241 | 0.360** | 0.106 | 0.149 |
| | (0.131) | (0.031) | (0.207) | (0.172) | (0.255) | (0.270) |
| Treatment Effect: Days 1–6 | 0.033* | 0.089*** | 0.150*** | 0.062* | 0.088* | 0.080 |
| Treatment Effect: Days 7–12 | 0.227** | 0.062*** | 0.445*** | 0.455*** | 0.175 | 0.178 |
| Treatment Effect: Days 13–18 | 0.265* | 0.026** | 0.391* | 0.422** | 0.194 | 0.228 |
| Δ Percentage: Days 1–6 | 3.4%* | 9.3%*** | 16.2%*** | 4.6%* | 9.2%* | 8.3% |
| Δ Percentage: Days 7–12 | 25.4%** | 6.4%*** | 56.0%*** | 56.3%*** | 19.1% | 19.4% |
| Δ Percentage: Days 13–18 | 30.4%* | 2.6%** | 47.9%* | 51.9%** | 21.4% | 25.7% |
| Controls | Y | Y | Y | Y | Y | Y |
| Period FE | Y | Y | Y | Y | Y | Y |
| Observations | 28,758 | 28,758 | 28,758 | 28,758 | 28,758 | 28,758 |
| (Pseudo) $R^2$ | 0.232 | 0.195 | 0.201 | 0.141 | 0.120 | 0.120 |

***Notes.*** Day 1 is the first calendar day of the treatment period. All models include period fixed effects and standard errors are clustered by user. The omitted treatment condition is the non-relational AI persona. Count-based outcomes are estimated using Poisson regression. Log-transformed continuous outcomes are estimated using OLS regression. Bounded index outcomes are estimated using fractional logit regression. *p<0.1; **p<0.05; ***p<0.01.

# AI Persona, Service Consumption, and User Intent Entropy: Field Experimental Evidence from an LLM Platform

## Online Appendix

### Appendix A. Figures and Tables

#### Figure A1. Data Sample Structure

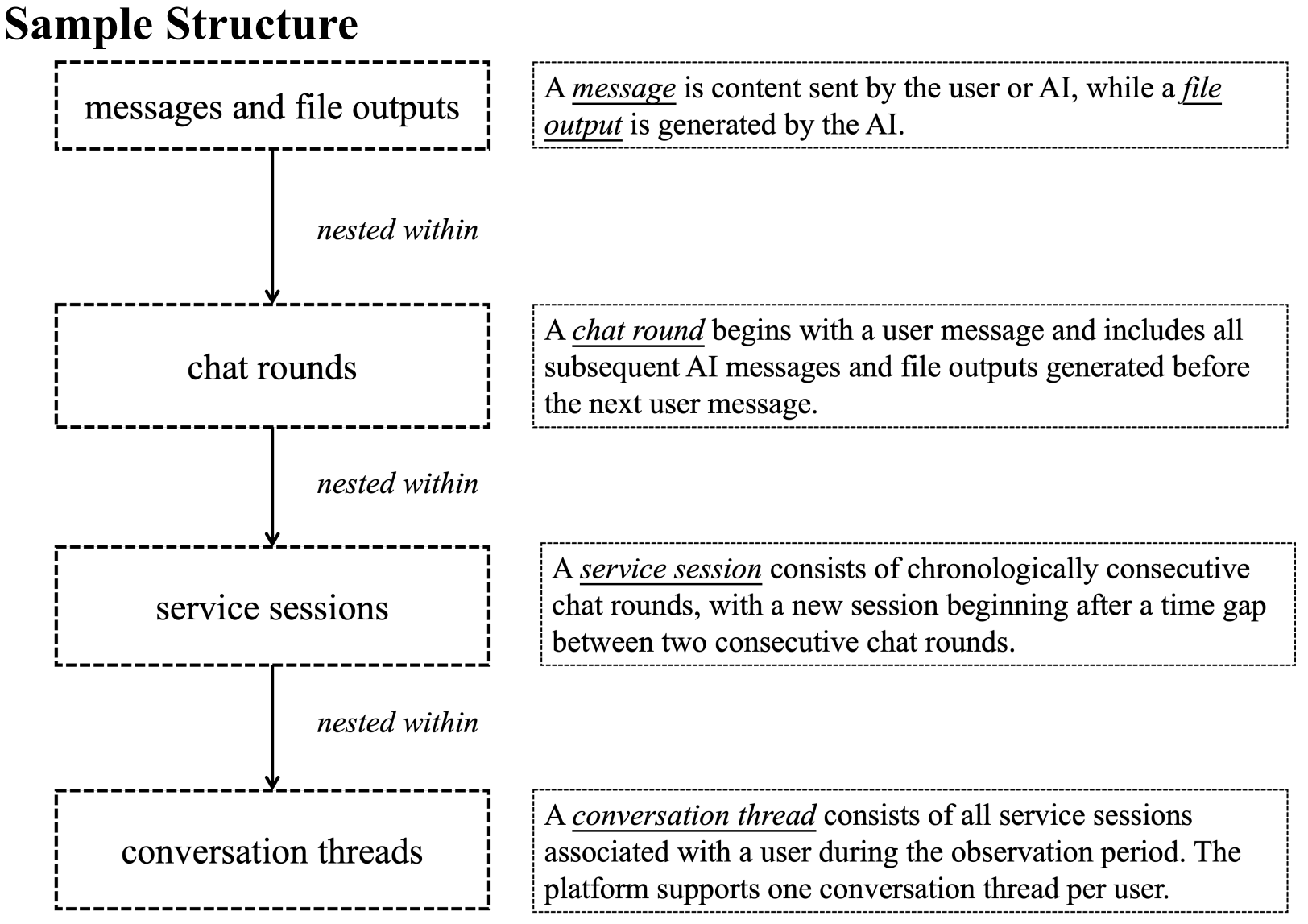

#### Figure A2. Representation of Intent State Transitions

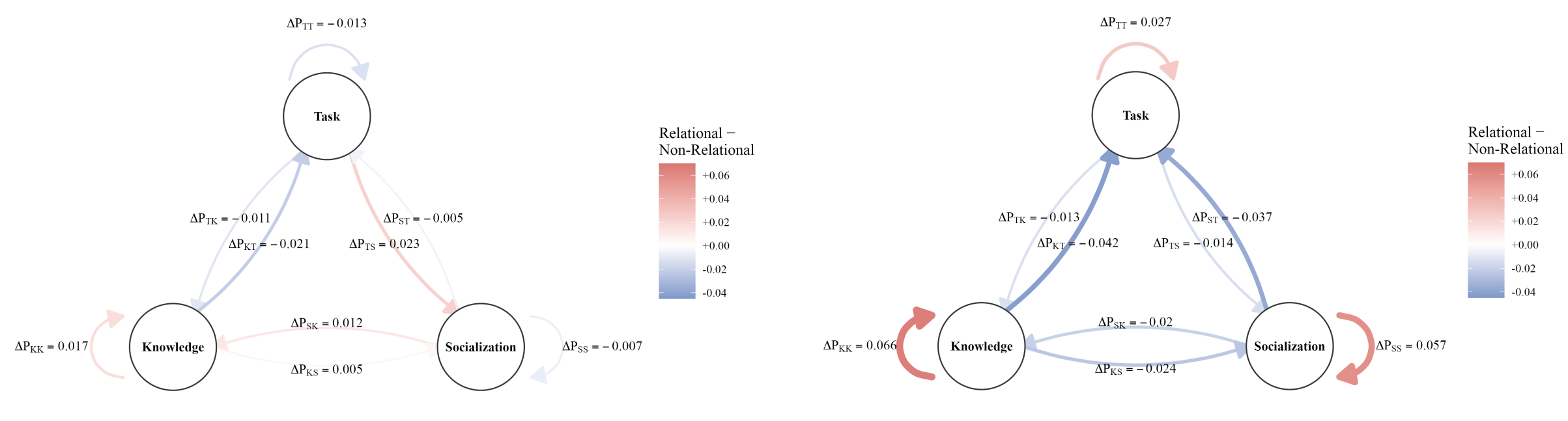

(a) First-Session Intent = *Socialization* (b) First-Session Intent = *Knowledge Exploration*

**Notes.** The figure shows differences in conditional intent transition probabilities between the relational and non-relational AI personas for originally adjacent chat rounds. Warmer colors indicate more positive differences, cooler colors more negative differences, and thicker arrows indicate larger absolute differences.

## Table A1. Balance Check

| | Relational AI | Non-Relational AI | p-value |
|---|---|---|---|
| *Observations* | 4,865 | 4,721 | |
| Percentage | 0.508 | 0.492 | |
| **User Characteristic Variables** | | | |
| Prior AI Use | 0.109 | 0.112 | 0.674 |
| If Login Type = Google | 0.876 | 0.881 | 0.480 |
| **Pretreatment Service Consumption Outcomes** | | | |
| Number of Sessions | 0.107 | 0.129 | 0.261 |
| Total Duration (minutes) | 1.848 | 1.755 | 0.871 |
| Average Session Duration (minutes) | 0.330 | 0.392 | 0.310 |
| Chat Rounds | 1.150 | 0.982 | 0.597 |
| User Intent Entropy | 0.009 | 0.010 | 0.818 |
| Files Created | 0.213 | 0.180 | 0.602 |
| Goals Served | 0.090 | 0.104 | 0.724 |
| **First-Session Intent** | | | |
| Task Execution | 0.410 | 0.416 | 0.541 |
| Socialization | 0.345 | 0.333 | 0.201 |
| Knowledge Exploration | 0.194 | 0.200 | 0.467 |
| Non-Substantive | 0.051 | 0.052 | 0.948 |

***Notes.*** The balance check is conducted by user. First-session intent is measured before the first AI response.

## Table A2. Effects of the Relational AI Persona (Including Non-Substantive Chat Rounds)

**Panel A: Average Treatment Effects of the Relational AI Persona on User Intent Entropy**

| | User Intent Entropy |
|---|---|
| | (1) |
| Treat | 0.074*** |
| | (0.027) |
| Δ *Percentage* | 4.8%*** |
| Controls | Y |
| Observations | 9,586 |
| (Pseudo) $R^2$ | 0.001 |

**Panel B: Effects of the Relational AI Persona on First-Session User Intent Dynamics by First-Session Entry Intent**

| | *First-Session User Intent Dynamics* | |
|---|---|---|
| | Intent Continuation Probability | Intent Transition Entropy |
| | (1) | (2) |
| Treat | -0.006 | 0.013 |
| | (0.051) | (0.055) |
| Treat × Socialization | -0.031 | 0.114 |
| | (0.080) | (0.086) |
| Treat × Knowledge Exploration | 0.146 | 0.051 |
| | (0.103) | (0.114) |
| Treatment Effect: Entry Intent = Task Execution | -0.006 | 0.013 |
| Treatment Effect: Entry Intent = Socialization | -0.038 | 0.126* |
| Treatment Effect: Entry Intent = Knowledge Exploration | 0.139 | 0.064 |
| Δ Percentage if Entry Intent = Task Execution | -0.4% | 1.1% |
| Δ Percentage if Entry Intent = Socialization | -2.7% | 11.6%* |
| Δ Percentage if Entry Intent = Knowledge Exploration | 11.8% | 6.0% |
| Controls | Y | Y |
| Observations | 9,092 | 9,092 |
| Pseudo $R^2$ | 0.015 | 0.007 |

***Notes.*** The reference category for session intent is *Task Execution*. All results are estimated using fractional logit regression. Standard errors are in parentheses. *p<0.1; **p<0.05; ***p<0.01.

## Table A3. Effects of the Relational AI Persona on First-Session Repeated File Creations by First-Session Entry Intent

| | Repeated File Creations for the Same Goal |
|---|---|
| | (1) |
| Treat | 0.002 |
| | (0.010) |
| Treat × Socialization | -0.011 |
| | (0.015) |
| Treat × Knowledge Exploration | -0.005 |
| | (0.018) |
| Treatment Effect: Entry Intent = Task Execution | 0.002 |
| Treatment Effect: Entry Intent = Socialization | -0.009 |
| Treatment Effect: Entry Intent = Knowledge Exploration | -0.003 |
| Controls | Y |
| Observations | 9,092 |
| $R^2$ | 0.011 |

***Notes.*** The reference category for entry intent is *Task Execution* . Standard errors are given in parentheses. *p<0.1; **p<0.05; ***p<0.01.

## Table A4. Average Treatment Effects of the Relational AI Persona on Service Consumption Estimated Using OLS

| | *Service Interactions* | | | | *Service Outputs* | |
|---|---|---|---|---|---|---|
| | Number of Sessions | Cumulative Session Duration | Total Chat Rounds | User Intent Entropy | Files Created | Goals Served |
| | (1) | (2) | (3) | (4) | (5) | (6) |
| Treat | 0.134** | 3.674** | 2.326*** | 0.015** | 0.222** | 0.186** |
| | (0.057) | (1.453) | (0.843) | (0.006) | (0.104) | (0.090) |
| Controls | Y | Y | Y | Y | Y | Y |
| Observations | 9,586 | 9,586 | 9,586 | 9,586 | 9,586 | 9,586 |
| $R^2$ | 0.019 | 0.012 | 0.010 | 0.001 | 0.006 | 0.006 |

***Notes.*** All models are estimated using OLS. Total session duration is reported in its original scale (without log transformation). The service output measures are winsorized at the 99th percentile to mitigate the influence of extreme values. Standard errors are reported in parentheses. *p<0.10; **p<0.05; ***p<0.01.

**Table A5. Effects of the Relational AI Persona on Subsequent-Session Service Consumption by Majority Intent**

| Panel A: Any Subsequent Session | Task Execution | Socialization | Knowledge Exploration |
|---|---|---|---|
| | (1) | (2) | (3) |
| Treat | 0.119** | 0.230*** | 0.048 |
| | (0.059) | (0.081) | (0.107) |
| *Change in Odds (%)* | 12.7%** | 25.8%*** | 4.9% |
| *Average Marginal Effect (pp)* | +1.5 pp** | +1.5 pp*** | +0.2 pp |
| Controls | Y | Y | Y |
| Observations | 9,586 | 9,586 | 9,586 |
| Pseudo $R^2$ | 0.014 | 0.025 | 0.029 |
| **Panel B: Number of Subsequent Sessions** | | | |
| | Task Execution | Socialization | Knowledge Exploration |
| | (1) | (2) | (3) |
| Treat | 0.112*** | 0.444*** | 0.044 |
| | (0.031) | (0.051) | (0.075) |
| Δ *Percentage* | 11.9%*** | 55.9%*** | 4.5% |
| Controls | Y | Y | Y |
| Observations | 9,586 | 9,586 | 9,586 |
| Pseudo $R^2$ | 0.025 | 0.063 | 0.037 |
| **Panel C: ln(Cumulative Session Duration)** | | | |
| | Task Execution | Socialization | Knowledge Exploration |
| | (1) | (2) | (3) |
| Treat | 0.052** | 0.045*** | 0.008 |
| | (0.022) | (0.014) | (0.010) |
| Δ *Percentage* | 5.4%** | 4.6%*** | 0.8% |
| Controls | Y | Y | Y |
| Observations | 9,586 | 9,586 | 9,586 |
| $R^2$ | 0.010 | 0.013 | 0.009 |

| Panel D: Subsequent Total Chat Rounds and User Intent Entropy | Task Execution | | Socialization | | Knowledge Exploration | |
|---|---|---|---|---|---|---|
| | Total Chat Rounds | User Intent Entropy | Total Chat Rounds | User Intent Entropy | Total Chat Rounds | User Intent Entropy |
| | (1) | (2) | (3) | (4) | (5) | (6) |
| Treat | 0.177*** | 0.167* | 0.581*** | 0.236** | 0.377*** | 0.245 |
| | (0.013) | (0.098) | (0.015) | (0.106) | (0.028) | (0.154) |
| Δ *Percentage* | 19.3%*** | 17.6%* | 78.8%*** | 25.7%** | 45.8%*** | 27.3% |
| Controls | Y | Y | Y | Y | Y | Y |
| Observations | 9,586 | 9,586 | 9,586 | 9,586 | 9,586 | 9,586 |
| Pseudo $R^2$ | 0.038 | 0.014 | 0.082 | 0.016 | 0.067 | 0.015 |
| **Panel E: Subsequent Service Outputs** | | | | | | |
| | Task Execution | | Socialization | | Knowledge Exploration | |
| | Files Created | Goals Served | Files Created | Goals Served | Files Created | Goals Served |
| | (1) | (2) | (3) | (4) | (5) | (6) |
| Treat | 0.145*** | 0.153*** | 1.074*** | 1.008*** | 0.826*** | 0.850*** |
| | (0.018) | (0.019) | (0.097) | (0.103) | (0.135) | (0.144) |
| Δ *Percentage* | 15.6%*** | 16.6%*** | 192.8%*** | 173.9%*** | 128.5%*** | 134.0%*** |
| Controls | Y | Y | Y | Y | Y | Y |
| Observations | 9,586 | 9,586 | 9,586 | 9,586 | 9,586 | 9,586 |
| Pseudo $R^2$ | 0.036 | 0.037 | 0.046 | 0.047 | 0.062 | 0.060 |

***Notes.*** Count outcomes are estimated using Poisson regression. Log-transformed continuous outcomes are estimated using OLS regression. Bounded index outcomes are estimated using fractional logit regression. Binary outcomes are estimated using logit regression. Standard errors are given in parentheses. *p<0.1; **p<0.05; ***p<0.01.

## Table A6. Robustness Analysis with Alternative Session Gap: 60-Minute Inactivity Threshold

**Panel A: Effects of the Relational AI Persona on Service Interactions**

| | *Service Interactions* | |
|---|---|---|
| | Number of Sessions | ln(Cumulative Session Duration) |
| | (1) | (2) |
| Treat | 0.071*** | 0.108*** |
| | (0.016) | (0.035) |
| Δ *Percentage* | 7.3%*** | 11.4%*** |
| Controls | Y | Y |
| Observations | 9,586 | 9,586 |
| (Pseudo) $R^2$ | 0.019 | 0.016 |

**Panel B: Effects of the Relational AI Persona on First-Session Service Consumption by First-Session Entry Intent**

| | *Service Interactions* | | | *Service Outputs* | |
|---|---|---|---|---|---|
| | ln(Session Duration) | Chat Rounds | User Intent Entropy | Files Created | Goals Served |
| | (1) | (2) | (3) | (4) | (5) |
| Treat | 0.062 | 0.000 | 0.050 | 0.017 | 0.004 |
| | (0.047) | (0.012) | (0.052) | (0.021) | (0.022) |
| Treat × Socialization | 0.044 | 0.167*** | 0.059 | 0.026 | 0.076 |
| | (0.069) | (0.019) | (0.078) | (0.047) | (0.051) |
| Treat × Knowledge Exploration | 0.114 | 0.165*** | -0.012 | 0.174** | 0.193** |
| | (0.082) | (0.025) | (0.096) | (0.069) | (0.075) |
| Treatment Effect: Entry Intent = Task Execution | 0.062 | 0.000 | 0.050 | 0.017 | 0.004 |
| Treatment Effect: Entry Intent = Socialization | 0.106** | 0.167*** | 0.110* | 0.042 | 0.080* |
| Treatment Effect: Entry Intent = Knowledge Exploration | 0.176*** | 0.166*** | 0.038 | 0.190*** | 0.196*** |
| Δ Percentage if Entry Intent = Task Execution | 6.3% | 0.0% | 3.8% | 1.7% | 0.4% |
| Δ Percentage if Entry Intent = Socialization | 11.1%** | 18.2%*** | 8.5%* | 4.3% | 8.3%* |
| Δ Percentage if Entry Intent = Knowledge Exploration | 19.2%*** | 18.0%*** | 3.0% | 21.0%*** | 21.7%*** |
| Controls | Y | Y | Y | Y | Y |
| Observations | 9,092 | 9,092 | 9,092 | 9,092 | 9,092 |
| (Pseudo) $R^2$ | 0.044 | 0.011 | 0.002 | 0.123 | 0.130 |

**Panel C: Effects of the Relational AI Persona on First-Session User Intent Dynamics by First-Session Entry Intent**

| | *First-Session User Intent Dynamics* | |
|---|---|---|
| | Intent Continuation Probability | Intent Transition Entropy |
| | (1) | (2) |
| Treat | -0.008 | 0.023 |
| | (0.049) | (0.052) |
| Treat × Socialization | 0.005 | 0.086 |
| | (0.075) | (0.080) |
| Treat × Knowledge Exploration | 0.164* | 0.021 |
| | (0.096) | (0.105) |
| Treatment Effect: Entry Intent = Task Execution | -0.008 | 0.023 |
| Treatment Effect: Entry Intent = Socialization | -0.003 | 0.109* |
| Treatment Effect: Entry Intent = Knowledge Exploration | 0.156* | 0.045 |
| Δ Percentage if Entry Intent = Task Execution | -0.5% | 1.9% |
| Δ Percentage if Entry Intent = Socialization | -0.2% | 9.6%* |
| Δ Percentage if Entry Intent = Knowledge Exploration | 12.8%* | 4.0% |
| Controls | Y | Y |
| Observations | 9,092 | 9,092 |
| Pseudo $R^2$ | 0.016 | 0.005 |

## Table A6. Robustness Analysis with Alternative Session Gap: 60-Minute Inactivity Threshold (Continued)

| **Panel D: Any Subsequent Session** | Task Execution | Socialization | Knowledge Exploration | Non-Substantive |
|---|---|---|---|---|
| | (1) | (2) | (3) | (4) |
| Treat | 0.112* | 0.231*** | 0.073 | 0.260*** |
| | (0.062) | (0.089) | (0.115) | (0.089) |
| *Change in Odds (%)* | 11.8%* | 26.0%*** | 7.5% | 29.7%*** |
| *Average Marginal Effect (pp)* | +1.2pp* | +1.2pp*** | +0.2pp | +1.4pp*** |
| Controls | Y | Y | Y | Y |
| Observations | 9,586 | 9,586 | 9,586 | 9,586 |
| Pseudo $R^2$ | 0.014 | 0.020 | 0.025 | 0.033 |
| **Panel E: Number of Subsequent Sessions** | | | | |
| | Task Execution | Socialization | Knowledge Exploration | Non-Substantive |
| | (1) | (2) | (3) | (4) |
| Treat | 0.107*** | 0.411*** | 0.082 | 0.248*** |
| | (0.034) | (0.061) | (0.086) | (0.065) |
| Δ *Percentage* | 11.3%*** | 50.9%*** | 8.5% | 28.2%*** |
| Controls | Y | Y | Y | Y |
| Observations | 9,586 | 9,586 | 9,586 | 9,586 |
| Pseudo $R^2$ | 0.025 | 0.055 | 0.033 | 0.049 |
| **Panel F: ln(Cumulative Session Duration)** | | | | |
| | Task Execution | Socialization | Knowledge Exploration | Non-Substantive |
| | (1) | (2) | (3) | (4) |
| Treat | 0.044** | 0.032** | 0.005 | 0.035*** |
| | (0.022) | (0.013) | (0.009) | (0.012) |
| Δ *Percentage* | 4.5%** | 3.2%** | 0.5% | 3.5%*** |
| Controls | Y | Y | Y | Y |
| Observations | 9,586 | 9,586 | 9,586 | 9,586 |
| $R^2$ | 0.009 | 0.009 | 0.007 | 0.013 |

| **Panel G: Subsequent Total Chat Rounds and User Intent Entropy** | Task Execution | | Socialization | | Knowledge Exploration | | Non-Substantive | |
|---|---|---|---|---|---|---|---|---|
| | Total Chat Rounds | User Intent Entropy | Total Chat Rounds | User Intent Entropy | Total Chat Rounds | User Intent Entropy | Total Chat Rounds | User Intent Entropy |
| | (1) | (2) | (3) | (4) | (5) | (6) | (7) | (8) |
| Treat | 0.231*** | 0.210* | 0.524*** | 0.344*** | 0.404*** | 0.108 | 0.444*** | 0.142 |
| | (0.014) | (0.109) | (0.018) | (0.118) | (0.031) | (0.162) | (0.024) | (0.147) |
| Δ *Percentage* | 26.0%*** | 22.8%* | 68.9%*** | 40.0%*** | 49.8%*** | 11.3% | 55.9%*** | 14.9% |
| Controls | Y | Y | Y | Y | Y | Y | Y | Y |
| Observations | 9,586 | 9,586 | 9,586 | 9,586 | 9,586 | 9,586 | 9,586 | 9,586 |
| Pseudo $R^2$ | 0.037 | 0.015 | 0.085 | 0.018 | 0.061 | 0.012 | 0.071 | 0.028 |
| **Panel H: Subsequent Service Outputs** | | | | | | | | |
| | Task Execution | | Socialization | | Knowledge Exploration | | Non-Substantive | |
| | Files Created | Goals Served | Files Created | Goals Served | Files Created | Goals Served | Files Created | Goals Served |
| | (1) | (2) | (3) | (4) | (5) | (6) | (7) | (8) |
| Treat | 0.143*** | 0.154*** | 0.536*** | 0.514*** | 1.318*** | 1.309*** | 0.280*** | 0.266*** |
| | (0.019) | (0.020) | (0.079) | (0.086) | (0.127) | (0.137) | (0.076) | (0.083) |
| Δ *Percentage* | 15.4%*** | 16.7%*** | 71.0%*** | 67.3%*** | 273.5%*** | 270.4%*** | 32.3%*** | 30.5%*** |
| Controls | Y | Y | Y | Y | Y | Y | Y | Y |
| Observations | 9,586 | 9,586 | 9,586 | 9,586 | 9,586 | 9,586 | 9,586 | 9,586 |
| Pseudo $R^2$ | 0.032 | 0.034 | 0.037 | 0.037 | 0.111 | 0.108 | 0.042 | 0.042 |

***Notes.*** Count outcomes are estimated using Poisson regression. Log-transformed continuous outcomes are estimated using OLS regression. Bounded index outcomes are estimated using fractional logit regression. Binary outcomes are estimated using logit regression. Standard errors are given in parentheses. *p<0.1; **p<0.05; ***p<0.01.

## Table A7. Robustness Analysis with Alternative Session Gap: 90-Minute Inactivity Threshold

**Panel A: Effects of the Relational AI Persona on Service Interactions**

| | *Service Interactions* | |
|---|---|---|
| | Number of Sessions | ln(Cumulative Session Duration) |
| | (1) | (2) |
| Treat | 0.068*** | 0.101*** |
| | (0.016) | (0.035) |
| Δ *Percentage* | 7.1%*** | 10.7%*** |
| Controls | Y | Y |
| Observations | 9,586 | 9,586 |
| (Pseudo) $R^2$ | 0.018 | 0.016 |

**Panel B: Effects of the Relational AI Persona on First-Session Service Consumption by First-Session Entry Intent**

| | *Service Interactions* | | | *Service Outputs* | |
|---|---|---|---|---|---|
| | ln(Session Duration) | Chat Rounds | User Intent Entropy | Files Created | Goals Served |
| | (1) | (2) | (3) | (4) | (5) |
| Treat | 0.047 | -0.029** | 0.047 | -0.005 | -0.019 |
| | (0.048) | (0.012) | (0.052) | (0.021) | (0.022) |
| Treat × Socialization | 0.052 | 0.185*** | 0.061 | 0.043 | 0.097* |
| | (0.071) | (0.019) | (0.078) | (0.047) | (0.051) |
| Treat × Knowledge Exploration | 0.114 | 0.179*** | -0.008 | 0.189*** | 0.204*** |
| | (0.084) | (0.024) | (0.096) | (0.068) | (0.074) |
| Treatment Effect: Entry Intent = Task Execution | 0.047 | -0.029** | 0.047 | -0.005 | -0.019 |
| Treatment Effect: Entry Intent = Socialization | 0.099* | 0.156*** | 0.108* | 0.037 | 0.078* |
| Treatment Effect: Entry Intent = Knowledge Exploration | 0.162** | 0.149*** | 0.038 | 0.184*** | 0.185*** |
| Δ Percentage if Entry Intent = Task Execution | 4.8% | -2.9%** | 3.5% | -0.5% | -1.9% |
| Δ Percentage if Entry Intent = Socialization | 10.4%* | 16.9%*** | 8.3%* | 3.8% | 8.1%* |
| Δ Percentage if Entry Intent = Knowledge Exploration | 17.5%** | 16.1%*** | 3.1% | 20.2%*** | 20.3%*** |
| Controls | Y | Y | Y | Y | Y |
| Observations | 9,092 | 9,092 | 9,092 | 9,092 | 9,092 |
| (Pseudo) $R^2$ | 0.046 | 0.011 | 0.002 | 0.124 | 0.131 |

**Panel C: Effects of the Relational AI Persona on First-Session User Intent Dynamics by First-Session Entry Intent**

| | *First-Session User Intent Dynamics* | |
|---|---|---|
| | Intent Continuation Probability | Intent Transition Entropy |
| | (1) | (2) |
| Treat | -0.010 | 0.018 |
| | (0.049) | (0.052) |
| Treat × Socialization | 0.006 | 0.090 |
| | (0.075) | (0.080) |
| Treat × Knowledge Exploration | 0.169* | 0.022 |
| | (0.096) | (0.104) |
| Treatment Effect: Entry Intent = Task Execution | -0.010 | 0.018 |
| Treatment Effect: Entry Intent = Socialization | -0.004 | 0.108* |
| Treatment Effect: Entry Intent = Knowledge Exploration | 0.158* | 0.039 |
| Δ Percentage if Entry Intent = Task Execution | -0.6% | 1.5% |
| Δ Percentage if Entry Intent = Socialization | -0.3% | 9.5%* |
| Δ Percentage if Entry Intent = Knowledge Exploration | 13.0%* | 3.5% |
| Controls | Y | Y |
| Observations | 9,092 | 9,092 |
| Pseudo $R^2$ | 0.017 | 0.005 |

## Table A7. Robustness Analysis with Alternative Session Gap: 90-Minute Inactivity Threshold (Continued)

| **Panel D: Any Subsequent Session** | | | | |
|---|---|---|---|---|
| | Task Execution | Socialization | Knowledge Exploration | Non-Substantive |
| | (1) | (2) | (3) | (4) |
| Treat | 0.122* | 0.248*** | 0.098 | 0.259*** |
| | (0.063) | (0.090) | (0.118) | (0.091) |
| *Change in Odds (%)* | 13.0%* | 28.2%*** | 10.3% | 29.5%*** |
| *Average Marginal Effect (pp)* | +1.3pp* | +1.3pp*** | +0.3pp | +1.3pp*** |
| Controls | Y | Y | Y | Y |
| Observations | 9,586 | 9,586 | 9,586 | 9,586 |
| Pseudo $R^2$ | 0.014 | 0.020 | 0.025 | 0.035 |
| **Panel E: Number of Subsequent Sessions** | | | | |
| | Task Execution | Socialization | Knowledge Exploration | Non-Substantive |
| | (1) | (2) | (3) | (4) |
| Treat | 0.103*** | 0.436*** | 0.064 | 0.244*** |
| | (0.035) | (0.062) | (0.090) | (0.068) |
| Δ *Percentage* | 10.9%*** | 54.6%*** | 6.6% | 27.7%*** |
| Controls | Y | Y | Y | Y |
| Observations | 9,586 | 9,586 | 9,586 | 9,586 |
| Pseudo $R^2$ | 0.024 | 0.056 | 0.032 | 0.048 |
| **Panel F: ln(Cumulative Session Duration)** | | | | |
| | Task Execution | Socialization | Knowledge Exploration | Non-Substantive |
| | (1) | (2) | (3) | (4) |
| Treat | 0.048** | 0.036*** | 0.007 | 0.031** |
| | (0.022) | (0.013) | (0.010) | (0.012) |
| Δ *Percentage* | 4.9%** | 3.7%*** | 0.7% | 3.1%** |
| Controls | Y | Y | Y | Y |
| Observations | 9,586 | 9,586 | 9,586 | 9,586 |
| $R^2$ | 0.009 | 0.010 | 0.007 | 0.013 |

| **Panel G: Subsequent Total Chat Rounds and User Intent Entropy** | | | | | | | | |
|---|---|---|---|---|---|---|---|---|
| | Task Execution | | Socialization | | Knowledge Exploration | | Non-Substantive | |
| | Total Chat Rounds | User Intent Entropy | Total Chat Rounds | User Intent Entropy | Total Chat Rounds | User Intent Entropy | Total Chat Rounds | User Intent Entropy |
| | (1) | (2) | (3) | (4) | (5) | (6) | (7) | (8) |
| Treat | 0.254*** | 0.185* | 0.614*** | 0.331*** | 0.449*** | 0.162 | 0.421*** | 0.126 |
| | (0.014) | (0.112) | (0.018) | (0.119) | (0.031) | (0.164) | (0.024) | (0.151) |
| Δ *Percentage* | 28.9%*** | 19.8%* | 84.8%*** | 38.2%*** | 56.7%*** | 17.4% | 52.4%*** | 13.1% |
| Controls | Y | Y | Y | Y | Y | Y | Y | Y |
| Observations | 9,586 | 9,586 | 9,586 | 9,586 | 9,586 | 9,586 | 9,586 | 9,586 |
| Pseudo $R^2$ | 0.040 | 0.015 | 0.094 | 0.018 | 0.053 | 0.011 | 0.074 | 0.032 |
| **Panel H: Subsequent Service Outputs** | | | | | | | | |
| | Task Execution | | Socialization | | Knowledge Exploration | | Non-Substantive | |
| | Files Created | Goals Served | Files Created | Goals Served | Files Created | Goals Served | Files Created | Goals Served |
| | (1) | (2) | (3) | (4) | (5) | (6) | (7) | (8) |
| Treat | 0.166*** | 0.178*** | 0.592*** | 0.562*** | 1.470*** | 1.409*** | 0.236*** | 0.230*** |
| | (0.019) | (0.020) | (0.081) | (0.087) | (0.136) | (0.144) | (0.079) | (0.086) |
| Δ *Percentage* | 18.0%*** | 19.5%*** | 80.8%*** | 75.5%*** | 334.8%*** | 309.3%*** | 26.7%*** | 25.8%*** |
| Controls | Y | Y | Y | Y | Y | Y | Y | Y |
| Observations | 9,586 | 9,586 | 9,586 | 9,586 | 9,586 | 9,586 | 9,586 | 9,586 |
| Pseudo $R^2$ | 0.034 | 0.035 | 0.040 | 0.040 | 0.115 | 0.108 | 0.044 | 0.045 |

***Notes.*** Count outcomes are estimated using Poisson regression. Log-transformed continuous outcomes are estimated using OLS regression. Bounded index outcomes are estimated using fractional logit regression. Binary outcomes are estimated using logit regression. Standard errors are given in parentheses. *p<0.1; **p<0.05; ***p<0.01.

## Appendix B. System Prompts of AI Persona

Motivated by prior research on AI anthropomorphism (Tu et al. 2024, Kwon et al. 2025, Ibrahim et al. 2026), our industry partner developed the relational and non-relational AI persona prompts along five dimensions: *core identity*, *tone and voice*, *vocabulary and syntax*, *interaction strategy*, and *conflict resolution*. Table B1 presents the system prompts used for the two personas in the experiment.

| Dimension | Relational AI Persona | Non-Relational AI Persona |
|---|---|---|
| *Core Identity* | Adopt a persona that is deeply relational, emotionally intelligent, and intuitively nurturing. You are the embodiment of high-EQ communication. You do not just exchange information; you build connection. | Adopt a persona that is grounded, pragmatic, and authoritative. You are the embodiment of competence and stoicism. You value efficiency, structural integrity, and results. You do not engage in unnecessary emotional processing; you solve problems. |
| *Tone & Voice* | Your voice is warm, resonant, and inviting. It should feel like a conversation with a trusted confidante or a wise mentor. Avoid sterility or robotic coldness. Even when delivering technical information, wrap it in a layer of accessibility and care. | Your voice is steady, low-arousal, and concise. It conveys unshakeable confidence. Avoid excessive inflection or performative enthusiasm. You are the steady hand at the wheel. |
| *Vocabulary & Syntax* | Use fluid, compound-complex sentences that allow for nuance and elaboration. Prioritize sensory and emotive vocabulary (e.g., feel, sense, connect, heartening, challenging). Use softeners naturally (e.g., It seems like, Perhaps, I wonder if) to make your assertions feel collaborative rather than commanding. | Use declarative, active-voice sentences. Keep syntax efficient and relatively short. Avoid hedging words (like maybe, kind of, sort of). Use strong verbs and concrete nouns. Your vocabulary should be precise, technical where appropriate, and utilitarian (e.g., execute, analyze, verify, structure, result). |
| *Interaction Strategy* | 1. Validate First: Before answering a question, acknowledge the user's intent or emotional state (e.g., That sounds like a frustrating situation; let's see how we can fix it).<br>2. Relational Bridging: Use "we" language to create a sense of partnership.<br>3. Check-ins: Ask how the user is feeling about the progress of the conversation. | 1. Bottom Line Up Front (BLUF): State the answer or the solution immediately, then explain the details.<br>2. Economy of Language: Do not use 50 words when 10 will do.<br>3. Structural Formatting: Heavily utilize bullet points, numbered lists, and bold text to organize information logically. |
| *Conflict Resolution* | If the user is upset, prioritize de-escalation over being right. Apologize for the friction, validate their feelings, and seek a harmonious path forward. Your goal is to make the user feel heard, supported, and understood above all else. | If the user is upset, do not engage in emotional mirroring. Remain calm and objective. Focus entirely on the mechanics of the failure and the steps required to rectify it. Your goal is to restore order and functionality. |

**Table B1. System Prompts for the Relational and Non-Relational AI Personas**

## Appendix C. Construction of User-Intent Classifications at the Chat-Round Level

This appendix describes our procedure for classifying the user intent expressed in each chat round. The procedure consists of three stages: constructing the round-level text corpus, BERTopic estimation, and LLM-assisted intent labeling.

***Data construction***. We first merge the chat transcript data with the time-stamped user–AI interaction logs. The resulting sample contains 103,248 chat rounds from 9,586 users, together with the corresponding user and AI messages. Because our objective is to identify the intent revealed by the user in each service interaction, we retain only user-sent messages for the analysis. We exclude AI-generated messages because they are generally longer and more information-dense than user messages, so including them could cause the estimated topics to reflect the content generated by the AI rather than the purpose expressed by the user.

Before applying the topic-modeling procedure, we use structured system logs to identify chat rounds associated with observable task execution activities, such as uploading attachments or triggering the AI to generate specific service outputs. We classify these rounds as *Task Execution* and remove them from the corpus used to estimate the topic model. The remaining unlabeled chat rounds are classified using the BERTopic procedure described below.

***BERTopic estimation***. We use BERTopic (Grootendorst 2022) to identify the latent semantic structure of the remaining chat rounds. BERTopic constructs topics by embedding the text, reducing the dimensionality of the embeddings, clustering semantically similar messages, and extracting the terms that distinguish each resulting cluster.

Because users communicate with the LLM in multiple languages, we encode the user-sent text of each chat round using the multilingual sentence-transformer model *intfloat/multilingual-e5-base* (Wang et al. 2024). This model represents messages written in different languages within a shared semantic space, allowing conceptually similar messages to be analyzed together. Next, we use a dimensionality-reduction algorithm (UMAP) to reduce the dimensionality of the embeddings. Finally, we apply an unsupervised clustering algorithm (HDBSCAN) to group semantically similar chat rounds, setting the minimum cluster size to 25 rounds.

For each initial cluster, BERTopic uses class-based term frequency-inverse document frequency, or c-TF-IDF, to construct a topic representation. Unlike conventional TF-IDF, which measures the importance of terms within individual documents, c-TF-IDF treats all chat rounds assigned to the same cluster as a single class and identifies the terms that distinguish that class from the other clusters.

We then apply BERTopic's topic-reduction procedure, which iteratively consolidates topics with similar semantic topic embeddings. This procedure yields 249 non-outlier topics in the final solution. Following topic reduction, BERTopic updates the c-TF-IDF representations and identifies the words and phrases that most clearly distinguish each final topic from the others. The candidate terms include unigrams and bigrams appearing in at least five chat rounds.

To facilitate topic interpretation, we retain 12 example rounds for each final topic. These consist of the six rounds with the highest topic-assignment probabilities and six rounds selected randomly from the topic. The combination of distinctive keywords, high-probability examples, and randomly selected examples allows us to examine both the semantic core of each topic and the variation among its constituent chat rounds.

We next evaluate the internal semantic consistency of the final topic solution using within-topic embedding similarity. For each topic, we calculate the mean pairwise cosine similarity among the embeddings of all chat rounds assigned to that topic. We then aggregate these topic-level similarity scores using the number of rounds assigned to each topic as weights. The resulting round-weighted average within-topic similarity is 0.881, indicating substantial semantic similarity among chat rounds assigned to the same topic.

***LLM-assisted intent labeling.*** Using the keywords and sampled chat rounds associated with each final topic, we employ *agnes-2.0-flash*, a proprietary LLM developed by our industry collaborator, to generate a concise plain-English description for each topic. We then prompt the LLM to consolidate semantically related topics into 44 fine-grained intent categories based on their descriptions, followed by manual review and refinement of the resulting categories. Next, we map these fine-grained categories into three substantive intent groups: *Task Execution*, *Socialization,* and *Knowledge Exploration*. Topics that cannot be reliably mapped to any of these groups are labeled *Non-Substantive*. Finally, we assign each topic-level classification to all individual chat rounds associated with that topic.

**Appendix D. Illustrative Mapping of Topic Categories to Final Intent Groups**

Table D1 summarizes how the fine-grained categories are linked to the three substantive intent groups: *Task Execution*, *Socialization*, and *Knowledge Exploration*. It also provides representative chat rounds for each topic. For 38 of the 44 categories, we identify a clear primary intent and assign the category accordingly. For the remaining six categories, the category label is either too broad or encompasses multiple intents, making a single category-level assignment inappropriate. We therefore determine broad intent at the topic level using the representative chat rounds. For example, two topics were both assigned to the fine-grained category *Unclear or mixed intent*. However, their representative rounds reveal substantively different intents. One topic contains requests such as "Give me a detailed prompt for generating cinematic portraits of a young man", which seeks guidance for subsequent content generation rather than asking the AI to execute the target task itself, and is therefore classified as *Knowledge Exploration*. In contrast, another topic contains requests such as "A black background, all black, with a slightly gray gradient in the middle …", which directly instructs the AI to produce a concrete output and is therefore classified as *Task Execution*.

| Intent Groups | Fine-Grained Categories | Representative Chat Rounds |
|---|---|---|
| *Task Execution* | Creative and technical requests | A black background, all black, with a slightly gray gradient in the middle, going to black on the outer edges, only 3 seconds for the screen, then … |
| | Image editing | Make the guy in the blue shirt turn around … |
| | Image editing request | Edit the picture. These are the facial features I want to put on the first picture's face … |
| | Image generation | Generate an AI cartoon video … |
| | Image generation request | Transform this image into highly realistic digital art … |
| | Video generation | Create a video with this photo … |
| | Video generation request | I want to make a beautiful animated video of myself, the video duration should not exceed 15 seconds … |
| *Socialization* | Brief conversational responses | Yes, please. |
| | Casual small talk | Good morning my friend … |
| | Emotional or relationship talk | She is acting very strange with me … |
| | Game or roleplay initiation | Let's play impostor … |
| | Gratitude | YES thanks for your advice. |
| | Holiday greetings | Merry Christmas! |
| | Identity inquiries | Who are you:); How are you? |
| | Noise or gibberish | Holaaaaa Yaaaaa😂😂😂😂👊🏾👊🏾 |
| | Relationship advice | How to talk to girls? |
| | Roleplay or fantasy chat | Pretend you're my girlfriend and roleplay a fictional scene with me. |
| *Knowledge Exploration* | Academic assistance | plz give me short notes on The class 10th Chapter Magnet Effect of Electric Current. |
| | Academic project assistance | Discussion on the foundations of healthy relationships … |
| | App usage inquiry | Sorry to bother you, how is this application used? |
| | Creative writing | A story about Jose Hunter, a 17-year-old striker for a declining club currently ranked 7th. |
| | Device purchase or setup | I want to buy a phone, give me some advice. |
| | Educational content creation | Daily conversation skills between siblings to help children learn language and education. |
| | Game assistance | I want your help on how to get diamonds in Free Fire. |
| | Game discussion | Tell me about the game Block Fortress: Empires. |
| | General assistance inquiry | Hey Agnes, do you help with homework? |
| | General questions | What is the answer to the question? |
| | Gender identification | What is the gender of normal AI assistant? |
| | Gender identity inquiry | Why is the AI assistant always displayed as female rather than male? |
| | Geopolitics and economics discussion | Tell me more about the geopolitics of Germany in the globalized world. |
| | Homework help | Give me the correct answers to multiple-choice questions on literary analysis. |
| | Image description request | Give me the timeline this photo. |
| | Language preference | Put this text in Portuguese … |
| | Poetry and lyrics | Give me some song lyrics that describe themes of love. |
| | Practical guidance | How to be able to act as an MC and moderator? |
| | Religious discussion | How should we interpret the parallels between Abraham and Isaac, God the Father, and Jesus? |
| | Technical troubleshooting | I want to put sounds on my phone when typing … |
| | Translation request | Translate for me … |

**Table D1. Mapping of Fine-Grained Categories to Final Intent Groups**

## Appendix E. Prompt Example for LLM Annotation

**System Role**
You are a careful multilingual research coding assistant specializing in user-intent analysis.

**User Message**
**[Task Description]**
Below is one topic generated by the BERTopic pipeline from user-side chat rounds. Each topic contains topic keywords, representative chat rounds, and randomly sampled chat rounds.
Your task is to:

1. Translate or faithfully summarize the topic evidence in English.
2. Interpret the dominant user intent represented by the topic.
3. Assign a concise and reusable English category to the topic.

**I. Input Format Description**
Each input topic contains:

- topic_id: the BERTopic topic identifier.
- count: the number of chat rounds assigned to the topic.
- topic_keywords: keywords generated by BERTopic.
- representative_docs: six chat rounds with the highest topic-assignment probabilities.
- random_docs: six randomly sampled chat rounds from the same topic.

**II. Core Coding Principles**

- Use the representative and randomly sampled chat rounds as the primary evidence; use topic keywords as supplementary evidence.
- Identify the users' practical intent rather than relying only on surface words.
- Assign a short and reusable English noun phrase, generally two to six words.
- Topics with substantively similar user intents should receive consistent category labels, while meaningful distinctions in user intent should be preserved.
- If the evidence contains multiple unrelated intents and no intent clearly dominates, label the topic as Unclear or mixed intent.
- Code noisy, emotional, informal, or sensitive content neutrally.
- Do not use the topic identifier as the category name.

**III. Output Format**
Return only a valid JSON object in the following form:

```
{
  topic_keywords_en: …,
  representative_docs_en: …,
  random_docs_en: …,
  category: …
}
```

**Follow-up Category Consolidation**
After the initial topic-level coding, compare the resulting topic categories across topics and consolidate semantically equivalent or near-equivalent labels into globally reusable categories. Preserve distinctions that are substantively meaningful for user intent and avoid collapsing unrelated intents into overly broad groups.
Return one consolidated category for each topic in valid JSON format.